\documentclass[hidelinks,11pt]{article}
\usepackage[pdftex]{color, graphicx}
\usepackage{amsmath, amsfonts, amssymb, mathrsfs, mathtools}
\usepackage{dcolumn}
\usepackage{natbib}
\usepackage{caption}
\usepackage{listings}
\usepackage{subcaption}
\usepackage{algpseudocode}
\usepackage[]{algorithm}
\usepackage{algorithmicx}
\usepackage{stackengine}
\usepackage{hyperref}
\usepackage{diagbox}

\bibpunct{(}{)}{;}{a}{}{,}
\newcommand{\x}{\mathbf{x}}
\newcommand{\X}{\mathbf{X}}
\newcommand{\Y}{\mathbf{Y}}
\newcommand{\K}{\mathbf{K}}
\newcommand{\kk}{\mathbf{k}}
\newcommand{\GG}{\mathbf{G}}
\newcommand{\g}{\mathbf{g}}
\newcommand{\Q}{\mathbf{Q}}
\newcommand{\W}{\mathbf{W}}

\newcommand{\I}{\mathbb{I}}
\newcommand{\OO}{\mathcal{O}}
\newcommand{\Sig}{\mathbf{\Sigma}}
\algrenewcommand{\algorithmiccomment}[1]{\hskip3mm\#\# #1}
\algnewcommand{\LeftComment}[1]{\Statex \#\# #1}

\begin{document}

\title{Locally induced Gaussian processes for \\large-scale simulation experiments%\thanks{DAC and RBG recognize support from National Science Foundation (NSF) grant DMS-1821258.}
}
\author{
	D.~Austin Cole\thanks{Corresponding author: \href{mailto:austin.cole8@vt.edu}{\tt austin.cole8@vt.edu}} \footnotemark[2]
	\and 
	Ryan Christianson\thanks{Department of Statistics, Virginia Tech, Blacksburg, VA}
	\and 
	Robert B.~Gramacy\footnotemark[2]
}

\maketitle

\begin{abstract} Gaussian processes (GPs) serve as flexible surrogates for
complex surfaces, but buckle under the cubic cost of matrix decompositions
with big training data sizes. Geospatial and machine learning communities
suggest pseudo-inputs, or inducing points, as one strategy to obtain an
approximation easing that computational burden.  However, we show how
placement of inducing points and their multitude can be thwarted by
pathologies, especially in large-scale dynamic response surface modeling
tasks.  As remedy, we suggest porting the inducing point idea, which is
usually applied globally, over to a more local context where selection is both
easier and faster. In this way, our proposed methodology hybridizes global
inducing point and data subset-based local GP approximation.  A cascade of
strategies for planning the selection of local inducing points is provided,
and comparisons are drawn to related methodology with emphasis on computer
surrogate modeling applications. We show that local inducing points extend
their global and data-subset component parts on the accuracy--computational
efficiency frontier.  Illustrative examples are provided on benchmark data and
a large-scale real-simulation satellite drag interpolation problem.
%\keywords{inducing points \and design \and surrogate \and approximation \and kriging \and  emulator}

\end{abstract}

\noindent

%\bigskip
%\noindent 
%\textbf{Declarations:}\\
%{\em Conflict of Interest}: The authors declare that they have no conflict of interest.\\
%{\em Data}: Robot arm SARCOS data \url{http://www.gaussianprocess.org/gpml/data/};
%Satdrag example: \url{https://bitbucket.org/gramacylab/tpm/src}.\\
%{\em Code Availability}:  {\sf R} code supporting all examples can be found in the
%Git repository \url{https://bitbucket.org/gramacylab/lagp/src/master/R/inducing/}.
%\pagebreak

\section{Introduction}
\label{sec:intro}

Advancements and expansion of access to supercomputing, algorithms for finite
element analysis, particle transport and agent-based modeling, combine in
modern times to produce simulation data of an unprecedented magnitude.  Yet as
modeling fidelity and configuration spaces continue to grow, coverage of
representative cases is still sparse. Gaussian process (GP) regression is a
common choice to fill in those gaps, emulating or serving as a surrogate for
the data-generating mechanism. GP surrogates excel at downstream tasks from
optimization to sensitivity analysis due to their out-of-sample predictive
accuracy and uncertainty quantification (UQ) capability, and ability to
interpolate the response when simulations are deterministic.  For a review of
computer experiments and surrogate modeling see \citet{Santer2018} or
\cite{gramacy2020surrogates}.

However, GP inference and prediction calculations scale poorly for large data
sets. GPs involve working with a multivariate normal (MVN) distribution whose
dimension matches the training data $(\X_N, \Y_N)$ size, $N$. Matrix
decomposition for covariance determinant and inverses is cubic in $N$. In
practice, this means limiting $N$ to the thousands -- small by modern
standards.

\sloppy Work from across disciplines where GPs play a fundamental role
(machine learning, geostatistics, computer experiments) targets remedies
through various approximations. Some methods induce sparsity in the covariance
\citep{titsias2009variational, aune2014parameter, Wilson2015, Gardner2018,
Pleiss2018, solin2020hilbert} or precision matrix \citep{Datta2016,
katzfuss2021}. Others propose divvying up the design space \citep{Kim2005,
Gramacy2008} and constructing multiple GPs by divide-and-conquer.
Partitioning offers the potential for parallelized multicore computation,
productively engaging untapped resources.  It also induces statistical
independence which can enhance flexibility when response surfaces have regime
changes or exhibit other nonstationary behavior.

\sloppy One framework, developed separately as {\em pseudo-inputs} in machine learning
\citep[e.g.,][]{Snelson2006} and {\em predictive processes} in geostatistics
\citep[e.g.,][]{Banerjee2008}, offers a low-rank approximation.  Together,
these two ideas are more recently referred to as {\em inducing point} methods.
Rather than measuring covariances between all pairs of $N$ training data
points directly, a smaller reference set $\bar{\X}_M$ of $M \ll N$ inducing
points or ``knots'' is used. Woodbury matrix identities make decompositions
cubic in $M$, a potentially dramatic savings. While space-filling
work well, optimizing the multitude $M$ and location of knots is fraught with challenges
\citep[e.g.,][]{Garton2020}.

One thing that sets surrogate modeling of computer simulations apart from
machine learning and geostats applications of GPs -- beside time being of the
essence -- is an all-but-total emphasis on prediction and UQ above other
inferential tasks. This opens up new opportunities for computational and
statistical economies by taking a {\em transductive}
approach to learning \citep{vapnik2013nature}: let the testing data dictate
how training is done.  Accurate, approximate GP prediction at an input $\x^\star$
can be based on a subset of data nearby $\x^\star$, leading to the so-called
local approximate GP \citep[LAGP;][]{Gramacy2015}.  Small data subsets $n \ll
N$ mean faster matrix decomposition, and potential for embarrassingly
parallel implementation \citep{gramacy2014massively}, through an infinite
divide-and-conquer/partition scheme.

The best sub-designs for predicting at $\x^\star$ depend on the training data
$\X_n(\x^\star) \subset \X_N$ nearby $\x^\star$. Those which are the very
closest -- a nearest neighbor (NN) subset
-- may not be ideal for all predictive goals, such as minimizing mean-squared
error
\citep[MSE;][]{vecchia1988estimation,stein2004approximating}.  Best results require
sequentially optimizing a criterion for each $\x^\star$ to greedily build
$\X_n(\x^\star)$.  Although speedy and vastly parallelizable, handling $N$
in the millions in a matter of minutes, it can still represent a
substantial computational effort, growing cubically with $n$ and
combinatorially in ${N\choose n}$ choices.  Authors have long opined that novel searches for
	each $\x^\star \in \X$ could be short-cut by learning some kind of
	re-locatable {\em template} of local sub-design characteristics
	\citep{gramacy2016speeding,sung2018exploiting}. However, a truly thrifty scheme
	has so far remained elusive.
	
	We believe a potential answer may lie in hybridizing inducing point and local GP schemes -- a variation on a recently popular theme of combining sparse
	GP methods with local models \citep{tan2016variational, liu2019understanding}.  The basic idea is as follows: 
	search locally for $m$ inducing points $\bar{\X}_m(\x^\star)$ in order to
	predict nearby $\x^\star$, specifically on a NN set $\X_n(\x^\star)$. Having
	$m \ll n \ll N$ leads to a manageable cascade of calculations.  We show how
	greedy optimization of $\bar{\X}_m(\x^\star)$, via a closed form weighted
	integrated MSE (wIMSE) criterion and gradients, avoids combinatorial
	sub-design search. Moreover, $\bar{\X}_m(\x^\star)$ can be used as a template,
	relocated anywhere for any $\x^\star$ without re-optimization.  In fact, we
	show that even locally space-filling schemes make for adequate templates in
	this setting.  The result is a {\em local inducing point GP (LIGP)}
	approximation which is nearly as accurate as LAGP, sometimes even more accurate, 
	and is faster.  Whereas LAGP was limited by small-$n$
	neighborhoods regardless of what the data prefer, we show
	that LIGP is not.  We explore neighborhoods more than double the size of
	LAGP and demonstrate accuracy improvements for commensurate computational
	effort.  This allows the user, for the first time, to fully explore the
	statistical--computational efficiency Pareto frontier in the context of local
	GP approximation.
	
	The remainder of the paper is organized as follows. Section
	\ref{sec:review} provides an overview of GP regression and various scalable
	models, including local and inducing points methods by way of motivating our
	hybrid approach. Section \ref{sec:liGP} describes the joining of local and
	inducing points methods comprising LIGP. We detail some refinements to LIGP,
	including local inducing point templates, in Section \ref{sec:refine}. Illustrative examples are provided
	throughout, however Section \ref{sec:examples} offers a systematic comparison
	of LIGP and LAGP variations to using both synthetic and real  benchmark
	examples. Section \ref{sec:discuss} concludes with a discussion.
	
	\section{Foundations in GP approximation}
	\label{sec:review}
	
	Here we highlight relevant surrogate modeling and scalable GP methods and
	provide motivation for a new criterion for placement of global and local
	inducing points.
	
	\subsection{Gaussian process regression}
	\label{sec:gprev} 
	
	Consider an unknown function $f: \X_N \subset \mathbb{R}^d \rightarrow
	\mathbb{R}$ for a set of $d$-dimensional design locations $\X_N=(\x_1,
	...,\x_N)^\top$ and corresponding observations $\Y_N=(y_1, ...,y_N)^\top$. GPs
	are common surrogates for such data \citep{Sacks1989}, especially as arising
	from deterministic computer simulations $f(\cdot)$, and boil down to
	placing an MVN prior on the observations $\Y_N$.  Gaussians are uniquely
	defined by a mean vector, which we take as zero for simplicity, and an $N
	\times N$ covariance matrix $\K_N$. The joint model for all responses is $\Y_N
	\sim \mathcal{N}_N(\mathbf{0},\nu (\K_N+\epsilon_k\I_N))$ where $\nu$ is a scale
	hyperparameter and $\K_N$ is comprised of entries based on a kernel
	$k_{\theta}(\x_i,\x_j)$. Jitter parameter $\epsilon_K$ is set as small as possible (for
	interpolating deterministic simulations) while maintaining well-conditioned
	positive-definite covariances \citep{Neal1998}, and $\I_N$ denotes an
	$N\times N$ identity matrix. Our presentation is agnostic to the
	choice of $k_\theta(\cdot,
	\cdot)$ except that it be based on inverse distances in the input space.  
	Our empirical work favors a squared exponential kernel
	with lengthscale hyperparameter $\theta$. 
	\begin{equation} 
		\label{eq:kernel}
		\K_N^{ij}=k_{\theta}(\x_i,\x_j)=\text{exp}\left\{
		-\frac{||\x_i-\x_j||^2}{\theta}\right\} 
	\end{equation} 
	Other common kernels include the Mat\'ern family
	\citep[][Section 5.3.3]{Stein2012,gramacy2020surrogates}.
	
	\sloppy Inference for unknown hyperparameters $(\theta, \nu)$ can proceed by maximum
	likelihood estimation through the log MVN pdf and its closed-form derivatives.  Some
	hyperparameters, like $\hat{\nu}=N^{-1}\Y_N^\top\K_N^{-1}\Y_N$, have tidy
	expressions conditional on others, like $\theta$, which must be optimized
	numerically.  Since MVN pdfs involve $|\K_N|$ and $\K_N^{-1}$, computation is
	on the order of $\OO(N^3)$, limiting training data sizes $N$ to the small
	thousands on most desktop machines.  In custom setups with highly distributed
	architectures,
	stochastic approximations based on linear
	conjugate gradients and Lanczos quadrature can push those boundaries
	\citep{ubaru2017fast,gardner2018gpytorch,wang2019exact}.
	
	For fixed hyperparameters $(\hat{\nu}, \hat{\theta})$ a predictive
	distribution for $Y(\x^\star)$ arises as standard MVN conditioning
	via an $(N+1)$-dimensional MVN for $(Y(\x^\star), \Y_N)$. The moments of
	that Gaussian distribution are:
	\begin{align} \label{eq:GPgen}
		\begin{split}
			\mu_N(\x^\star)&=\mathbb{E}(Y(\x^*)\mid\Y_N)
			 =\kk_N^\top(\x^\star)\K_N^{-1}\Y_N \\
			\sigma_N^2(\x^\star) &=\mathbb{V}\text{ar}(Y(\x^\star)\mid\Y_N) = \hat{\nu} \left(k_{\theta}(\x^\star,\x^\star)-\kk_N^\top(\x^\star)\K_N^{-1}\kk_N(\x^\star)\right),
		\end{split}
	\end{align} 
	where $\kk_N(\x^\star) = (k_{\theta}(\x^\star,\x_1), ...,
	k_{\theta}(\x^\star,\x_N))^{\top}$. These calculations are also
	in $\OO(N^3)$, although again linear
	algebra tricks can mitigate that to an extent.
	
	\subsection{Inducing points}
	\label{sec:inducing}
	
	A more direct approach to speedy GP approximation in the face of big $N$ is to
	impose a low-rank structure on covariance. The idea originated with local
	data subsets for splines \citep{Wahba1990, Poggio1990}, and later was applied to GPs
	\citep{Smola2001, csato2002sparse, Seeger2003}.
	\cite{Snelson2006} proposed that
	these reference locations not be restricted to a subset of the data. First
	attempts at a unifying perspective for sparse approximate GPs were made by
	\cite{Quinonero2005} and \citet[][Chapter 8]{Rasmussen2005}, with the former
	referring to these latent reference variables as {\em inducing inputs}. Outside
	of the machine learning community, \cite{Banerjee2008} applied similar
	techniques to develop {\em predictive processes}. Here we adopt a big-tent
	inducing points nomenclature.  
	
	Let $\bar{\X}_M=(\bar{\x}_1, \dots, \bar{\x}_M)^\top$ be $M$ inducing points
	in the same space as $\X_N$, but they need not coincide with any elements of
	$\X_N$. Notate $\K_M$ as a kernel matrix built from $\bar{\X}_M$ and
	$k_\theta(\cdot, \cdot)$, e.g., in \eqref{eq:kernel}; similarly, write
	$\kk_{NM}$ as cross evaluations of the kernel between $\X_N$ and $\bar{\X}_M$.
	Most variations on inducing point methods base GP approximations on the
	so-called Nystr\"{o}m approximation \citep{Williams2001}: $\K_N \approx
	\bar{\K} = \kk_{NM}\K_M^{-1}\kk_{NM}^\top$.  %See Figure \ref{fig:covar}. 
	Rather than calculate covariance between all pairs in ${\X}_N$,
	instead use $M \ll N$ references  $\bar{\X}_M$ to induce a similar structure
	$\bar{\K}$.
	
	\citet{Snelson2006} introduced a diagonal correction on the
	Nystr\"{o}m approximation
	\begin{equation} \label{eq:SPGP}
		\Sig_N^{(M)} =\nu(\bar{\K}+\epsilon_k\I_N)= \nu\Big(\kk_{NM}\K_M^{-1}\kk_{NM}^\top + \Lambda_N^{(M)} +\epsilon_K\I_N\Big)
	\end{equation}
	where $\Lambda_N^{(M)}=\text{Diag}\{\K_N-\kk_{NM}\K_M^{-1}\kk_{NM}^\top \}$. 
	This ensures that $\bar{\K}$ and $\K_N$ contain the same diagonal elements so
	that when $\bar{\X}_M\equiv \X_N$, $\Sig_N^{(M)}$ in \eqref{eq:SPGP} reduces
	to the standard GP covariance $\Sig_N = \nu (\K_N+\epsilon_k\I_N)$. Both approximations allow
	for decomposition of $\Sig_N^{(M)}$ through Woodbury matrix identities
	\citep{Harville2011}:
	\begin{align} 
			\Sig_N^{-1(M)}=&\nu^{-1}\Big(\Omega_N^{-1(M)}-\mathbf{\Gamma}_{NM}\Q_M^{-1(N)}\mathbf{\Gamma}_{NM}^{\top}\Big) \label{eq:Sigi} \\
			\log |\Sig_N^{(M)}|=&\log(\nu)+\log |\Q_M^{(N)}|- \log |\K_M| +\mathbf{1}_N^{\top}\text{log}(\Omega_N^{(M)})\mathbf{1}_N,  \nonumber
	\end{align}	
	where $\mathbf{\Gamma}_{NM}=\Omega_N^{-1(M)}\kk_{NM}$ and $\mathbf{1}_N$ is a vector of $N$ ones. Above, $\Q_M^{(N)}= \K_M +
	\kk_{NM}^{\top}\Omega_N^{-1(M)} \kk_{NM}+\epsilon_Q\I_M$ and
	$\Omega_N^{(M)} =\Lambda_N^{(M)} + \epsilon_K\I_N$. Since
	$\Omega_N^{(M)}$ is an $N\times N$ diagonal matrix and can be stored and
	manipulated as a vector, we elect to not embolden its notation like that of
	other matrices. 
	Hyperparameter inference is achieved by maximizing the logarithm of the MVN likelihood  $Y_N \sim
	\mathcal{N}\left(\mathbf{0},\nu(\kk_{NM}\K_M^{-1}\kk_{NM}^\top+\Omega^{(M)}_N )\right)$:
	\begin{align}
		\ell(\X,\Y,\bar{\X}_M,\nu, \theta, g) =& -\frac{N}{2}\log(2\pi)-\frac{1}{2}\log|\Sig_N|-\frac{1}{2}\Y_N^\top\Sig_N^{-1}\Y_N \label{eq:loglik}\\
		\propto & \mathrm{ const.} - N\log(\nu)-\log |\Q_M^{(N)}| + \log|\K_M| -  \mathbf{1}_N^{\top}\text{log}(\Omega_N^{(M)})\mathbf{1}_N  \nonumber\\
		&-\nu^{-1}\Y_N^\top\left(\Omega^{-1(M)}_N-\mathbf{\Gamma}_{NM}\Q_M^{-1(N)}\mathbf{\Gamma}_{NM}^{\top}\right)\Y_N. \nonumber
	\end{align}
	Differentiating Eq.~\eqref{eq:loglik} with respect to $\nu$ and solving yields the closed-form estimate
	\begin{equation}
		\hat{\nu}^{(N,M)} =  N^{-1}\Y_N^\top\left(\Omega^{-1(M)}_N-\mathbf{\Gamma}_{NM}\Q_M^{-1(N)}\mathbf{\Gamma}_{NM}^{\top}\right)\Y_N.\label{eq:nuhat} 
	\end{equation}
	There is not a similar closed-form solution for the lengthscale. Numerical solvers like $\tt{optim}$ in $\sf{R}$ can work with negative concentrated log-likelihood
	\begin{align} \label{eq:concentrate_ll}
			-\ell \; (\X,\Y,\bar{\X}_M, \theta)
			\propto& \; N\log\left(\Y_N^\top\left(\Omega^{-1(M)}_N-\mathbf{\Gamma}_{NM}\Q_M^{-1(N)}\mathbf{\Gamma}_{NM}^{\top}\right)\Y_N\right) \\
			&\;\; +\log |\Q_M^{(N)}| - \log|\K_M| +  \mathbf{1}_N^{\top}\text{log}(\Omega_N^{(M)})\mathbf{1}_N \nonumber
	\end{align}
	and closed form derivatives (not shown) to obtain $\hat{\theta}^{(N,M)}$.  In practice this works well because the surfaces are either convex in hyperparameters, or are nearly so.
	
	Analogues to  Eqs.~(\ref{eq:nuhat}--\ref{eq:concentrate_ll}) 
	reduce full rank prediction from $\mathcal{O}(N^3)$ down to $\OO(NM^2)$
	flops.  Following \eqref{eq:SPGP}, predictive equations are Gaussian with 
	\begin{align} 
		\label{eq:GPpred} 
		\mu_{M,N}(\x^\star)&=\kk_M^{\top}(\x^\star)\Q_M^{-1(N)}\mathbf{\Gamma}_{NM}^\top
		\Y_N \\ 
		\sigma_{M,N}^2(\x^\star)
		& =\nu \! \left(\!\K_{**}\!-\!\kk_M^{\top}(\x^\star)\left(\!\K_M^{-1}\!-\!\Q_M^{-1(N)}\!\right)\kk_M(\x^\star)\!\right),\nonumber
	\end{align} 
	\sloppy where $\kk_M(\x^\star)=k_{\theta}(\bar{\X}_M,\x^\star)$.
	When optimizing $\bar{\X}_M$ via log likelihood, the value
	$\Q_M^{-1(N)}\mathbf{\Gamma}_{NM}^\top\Y_N$ can be re-used from
	$\Sig_N^{-1(M)}\Y_N$. Thus, prediction requires only $\OO(M)$ and $\OO(M^2)$
	additional flops compared to $\OO(N)$ and $\OO(N^2)$ for a full GP model.
	
	\subsection{Optimal induction}
	\label{sec:opt}
	
	Suppose, for now, that the number of inducing points $M$ is fixed by
	computational limitations.  \cite{Snelson2006} suggested selecting
	locations $\bar{\X}_M$ through the marginal log-likelihood. Such a
	strategy is prone to overfitting \citep{bauer2016understanding}, while
	the Variational Free Energy (VFE) approximation---a lower bound on the marginal
	likelihood---is not \citep{titsias2009variational, Titsias2009,
	hoffman2013stochastic}. Yet even with VFE's variational construction of the
	likelihood, its optimization still requires a cubic cost on a
	highly-multimodal surface \citep{bauer2016understanding}, which we explore in
	Appendix \ref{app:imse}. This begs the question if likelihood optimization
	is worth it over relatively convenient space-filling options.
	
	Methods for ``choosing inputs'', known more widely as statistical design or {\em active learning}, have potential to reduce the cost of selecting inducing points. A slew of acquisition functions for greedy design point selection are available for such diverse goals as integral estimation \citep{fernandez2020adaptive, kanagawa2019convergence}, space-fillingness \citep{busby2009hierarchical, svendsen2020active}, and posterior density approximation \citep{wang2018adaptive}. Other variance-based \eqref{eq:GPgen}
	criteria may be appropriated for the selection
	of inducing points $\bar{\X}_M$ by routing through Eq.~(\ref{eq:GPpred})
	instead. Such criteria require quadratic computational cost and more squarely target
	predictive goals in surrogate modeling.	In particular, we consider integrated mean-squared
	error (IMSE) and its discretized analog Active Learning Cohn \citep[ALC]{Cohn1993} to select inducing
	points $\bar{\X}_M$, a design strategy not yet explored in the literature. See Appendix \ref{app:imse} for an overview of these variance-based criteria.	
	\begin{figure}[ht!]
		\centering
		\includegraphics[trim=5 5 25 55, clip, width=0.75\textwidth]{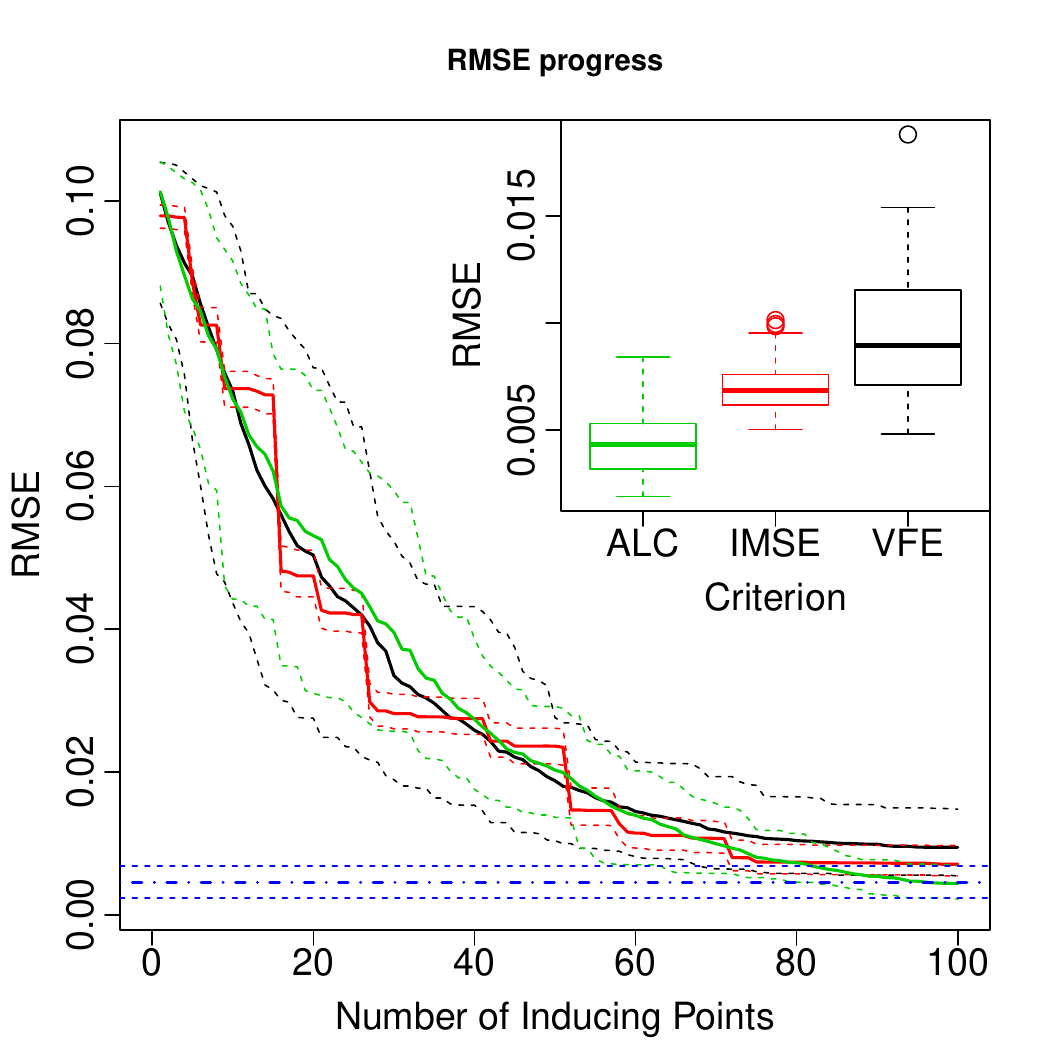}
		\caption{Approximate GP performance via RMSE and
			number of inducing points, $M$, compared to a full GP (blue).  Means (solid)
			and central 90\% intervals (dashed) arise from thirty replicates.
			Boxplots in the top right zoom in at
			$M=100$.}
		\label{fig:seqbakeoff}
	\end{figure}
	
	For a simple experiment, we sought to compare the predictive accuracy of sparse GP models with inducing points selected sequentially with VFE, IMSE, and ALC to a full GP. We generate data using $f(x_1, x_2)=x_1\exp\{-x_1^2-x_2^2\}$ for $x_1,x_2\in[-2,4]$. Figure \ref{fig:seqbakeoff} compares the three
	methods to themselves and to a full GP over $M=1,\dots,100$ tracking
	root MSE (RMSE) via Monte Carlo (MC)
	averaging over training $\X_N$ and testing $\mathcal{X}$ locations. To manage
	the computational cost of evaluating criteria on a dense grid, training data
	sizes were limited to $N=100$.\footnote{Ordinary IMSE was used, substituting
		inducing points in for design points, as described in \cite{hetGP2}.  Progress
		is blocky because individual inducing point additions do not substantial alter
		space-filling properties until most of a ``new row'' of sites are added in
		this 2d example.} Observe that all three methods offer a decent approximation
	to the full GP with close to 85 inducing points.  Zoomed boxplots (upper-right
	panel) show that ALC is consistently best.  If you know where you are going to
	be tested, you should ``design'' your $\bar{\X}_M$ to focus there. 
	If you do not, then you are (eventually) next-best by integrating over the input
	domain with IMSE. VFE performs worst because 
	likelihood is imperfectly aligned to the RMSE criteria.
	
	\subsection{Local approximate GPs}
	\label{sec:laGP}
	
	Rather than massage the GP framework to cope with the entire data set at once, e.g.,
	by working with a single global data subset, a {\em local approximate GP}
	\cite[LAGP;][]{Gramacy2015} considers disparate local data subsets depending
	on each of the predictive location(s) $\x^\star$ of interest. Such subsets can
	be much smaller because, under typical inverse-distance based correlation
	\eqref{eq:kernel}, training data inputs $\X_N$ far from each $\x^\star$ provide
	little added value to the underlying predictor.  Specifically, suppose that
	$(\X_n(\x^\star), \Y_n(\x^\star))$ represents an $n$-sized subset, or {\em
		neighborhood} of the training data nearby $\x^\star$, e.g., comprised of
	nearest neighbors (NNs). Then, given a suitable hyperparameterization,
	prediction could follow Eq.~\eqref{eq:GPgen} using $(\X_n(\x^\star),
	\Y_n(\x^\star))$ rather than the full $(\X_N, \Y_N)$. This can potentially provide drastic
	computational savings when $n \ll N$, even though the calculations would still
	be cubic in $n$.
	
	In this framework, the subset size $n$ and neighborhood $\X_n(\x^\star)$ must be determined.
	Because flops grow quickly with $n$, this value is usually fixed by
	computational limitations, just like the number of inducing points, $M$.  A
	default in the {\tt laGP} software \citep{laGP} is $n=50$,
	see Appendix \ref{app:nsize} for further discussion.  Fixing $n$, it turns out that NN subdesign, originally suggested
	by \cite{emery2009kriging} in a 2d geostatistics setting, is sub-optimal by
	several criteria \citep{vecchia1988estimation,Stein2012}. However,
	exhaustively searching among all ${N \choose n}$ alternatives for each
	$\x^\star$ is combinatorially infeasible.  \citeauthor{Gramacy2015} showed
	that greedy neighborhood selection via ALC approximately minimizes a
	MSE criteria common in surrogate modeling settings. 
	Specifically, choose a singleton reference set $\mathcal{X} = \{\x^\star\}$,
	with $\sigma_{\mathrm{new}}^2(\cdot) = \sigma_{n+1}^2(x)$ derived from
	$(\X_n(\x^\star), \Y_n(\x^\star))$ and select among $\x_{n+1} \in \X_N
	\setminus \X_n(\x^\star)$ candidates.\footnote{Here we are abusing notation a
		little to describe an inductive process $n \rightarrow n+1$ and referring to
		$n$ as the final local design size as opposed to introducing a new iterator.}

	Care is taken to ensure computational demands in each update and ALC
	optimization do not exceed $\OO(n^2)$ so that the
	entire scheme's flops are not worse in order than using NNs (i.e., cubic in
	$n$).  For example, if
	$v_n(\x_{n+1})=\kk_{n+1}(\x_{n+1},\x_{n+1})-\kk_{n+1}^{\top}(\x_{n+1})\K_{n+1}^{-1}\kk_{n+1}(\x_{n+1})$
	represents the kernel portion of $\sigma_n^2(\x_{n+1})$, then the change
	\begin{align}  \label{eq:MSE}
		\begin{split}
		\Delta v_n(\x^\star)&=v_n(\x^\star)-v_{n+1}(\x^\star)\\
		& =
		\kk_n^{\top}(\x^\star)\GG_n(\x_{n+1})v_n(\x_{n+1})\kk_n(\x^\star) +2\kk_n^{\top}(\x^\star)\g_n(\x_{n+1})k_{\theta}(\x_{n+1},\x^\star)\\
		&\quad+k_{\theta}(\x_{n+1},\x^\star)^2/v_n(\x_{n+1})
		\end{split}
	\end{align}
	can be updated in $\OO(n^2)$ via partition inverse
	equations \citep{barnett:1979} using
	$\GG_j(\x_{n+1})=\g_n(\x_{n+1})\g_n^{\top}(\x_{n+1})$,
	$\g_n(\x_{n+1})=-\K_n^{-1}\kk_n(\x_{n+1})/v_n(\x_{n+1})$.
	
	Despite being massively parallelizable \citep{gramacy2014massively} for many
	$\x^\star$ and over candidates $\x_{n+1} \in \X_N \setminus  \X_n(\x^\star)$,
	further approximations are made in order to shortcut $\OO(N)$ subroutines in
	an $\OO(n^2)$ scanning over that set
	\citep{gramacy2016speeding,sung2018exploiting,sun2019emulating}.  Several
	groups of authors have suggested that it might be possible to design a
	``template'' sub-design that could be applied automatically, after simple
	shifting/scaling for each $\x^\star$, without exhaustive search of $\X_N
	\setminus \X_n(\x^\star)$.  Non-uniform global designs $\X_N$ render this a
	non-starter. Sparse design coverage in some regions, and dense in others,
	demands bespoke calculation in each $\x^\star$ instance.  Even with highly
	regular (e.g., gridded) global designs $\X_N$, local coverage can be irregular
	at the boundaries. 
	
	Local design topology is twinned with subset size, $n$. Accommodating
	wiggly test problems benefit with reactive dynamics
	offered by smaller $n$ is easy, because that means faster execution. But $n$
	much larger than the default of $n=50$ can be a deal-breaker on speed grounds
	regardless of accuracy boosts in less wiggly settings. 
	
	\section{Inducing point neighborhoods}
	\label{sec:liGP}
	
	Inducing points offer computational savings, but several
	drawbacks remain.  Predictive accuracy suffers when they are placed far from
	testing locations. Optimization by likelihood can perform worse than simple
	space-filling (Section \ref{sec:opt}).  Computational costs are still cubic in
	a big number, despite $M \ll N$ because you need enough $M$ to fill the input
	volume. Multi-processing parallel schemes via likelihood \citep{chen2013parallel} and stochastic variational inference \citep{hensman2013gaussian, hoang2015unifying, schurch2020recursive} offer limited respite because they operate on the full data.
	
	We thus propose a {\em locally induced GP (LIGP)} by hybridizing
	ordinary, ``global'' inducing point schemes with LAGP.  This brings knock-on
	benefits to the local data-subsetting world: speed-ups, selection of
	neighborhood size (larger for smoother processes), long-elusive
	template schemes (Section \ref{sec:refine}). LIGP operates similarly to LAGP
	via neighborhoods $\X_n(\x^\star) \subset \X_N$. If a greedy scheme like ALC is used to fill $\X_n(\x^\star)$, it would include an exhaustive search  on the order of $\OO(Nn^3)$.
	Instead we choose simple NN approach, incurring an amortized one-off $\OO(N \log N)$ cost. Effort is reallocated into choosing local inducing points
	$\bar{\X}_m(\x^\star)$ for $\X_n(\x^\star)$, which are free to take on any
	values, at cubic in $m$ cost. Our multiplicity notation is intended to convey
	$m \ll n \ll M \ll N$, although that hierarchy need not be strict. Small $m$
	allows wider local scope with bigger $n$ without a substantial computational
	hit.
	
	Algorithm \ref{alg:ligp_pred} outlines the LIGP prediction algorithm, which can be run independently for each $\x^\star \in \X^\star$. For each $\x^\star$, a local neighborhood $\X_n(\x^\star)$ is built from a NN subset of $\X_N$ followed by a set of inducing points $\bar{\X}_m(\x^\star)$. Various methods to select $\bar{\X}_m(\x^\star)$ are explored in the following sections.
		\begin{algorithm*} \caption{LIGP Prediction} \begin{algorithmic}[1] \Procedure
			{LIGP.pred}{$m$, $n$, $\X^\star$,$\X_N$,$\Y_N$, $\mathcal{X}$}
			\For{$i=1,\dots,N' = |\X^\star|$}\hfill
			\Comment{Each $\x^\star_l \in \X^\star$, potentially in parallel} \State
			$\{\bar{\X}_m, \X_n\} \leftarrow$ IP$(\dots)$ \hfill \Comment{Any of
				Algorithms \ref{alg:ip_wimse}--\ref{alg:qnorm_design}} \State $\Y_n \leftarrow
			Y(\X_n)$ \hfill \Comment{Extract from $\Y_N$ at neighborhood} \State
			$\hat{\nu},\hat{\theta} \leftarrow \mathrm{argmax}_{\nu,\theta} \; \text{LLik}
			(\nu, \theta, \X_n,\Y_n,\bar{\X}_m)$ \hfill \Comment{Local MLE,
				Eqs.~(\ref{eq:nuhat}--\ref{eq:concentrate_ll})} \State $\{\hat{\mu}^{(i)},
			\hat{\sigma}^{2(i)}\} \leftarrow $ {\sc GP.pred}($\x^\star_i\mid\X_n,\Y_n,
			\bar{\X}_m,\hat{\theta}, \hat{\nu}$) \hfill \Comment{Eq.~(\ref{eq:GPpred})}
			\EndFor \State \Return $\{\hat{\mu}^{(i)},\hat{\sigma}^{2(i)}\}_{i=1}^{N'}$
			\EndProcedure \end{algorithmic} \label{alg:ligp_pred} \end{algorithm*}
		
	\subsection{Sequential selection of local inducing points}
	\label{sec:seqIP}

	Changing focus to local neighborhoods $\X_n(\x^\star)$ warrants a second look
	at selection criteria for inducing points $\bar{\X}_m(\x^\star)$. Likelihoods
	here are a mismatch to surrogate modeling and machine learning predictive
	goals. Instead, we follow the LAGP format of greedy optimization via
	MSE.  Given the connection between inducing
	$\bar{\X}_m(\x^\star)$ and actual training locations $\X_n(\x^\star)$,
	emphasis on prediction at singleton $\x^\star$ has deleterious effects.
	We tried this: $\bar{\X}_m(\x^\star)$ ``pile up'' around $\x^\star$ leading
	to poor estimates of local lengthscale and curvature. Instead, we suggest a
	locally weighted IMSE criterion.
	
	Suppose we have $\bar{\X}_m(\x^\star)$ already and wish to choose the next
	inducing point $\bar{\x}_{m+1}(\x^\star)$.  Dependence on $\x^\star$ is
	implicit below, although we shall drop it from the expressions and simply
	write $\X_n$, $\bar{\X}_m$ and $\bar{\x}_{m+1}$, etc., in order to streamline
	the notation.  We presume that the study region is a hyperrectangle
	$\mathcal{X} = [a_k, b_k]_{k=1}^d$.  Rather than integrate uniformly over that
	domain, reproducing an ordinary global IMSE whose closed form slightly
	generalizes \cite{hetGP2}, we weight the calculation by proximity to the
	predictive location $\x^\star$.  Although this weighting scheme could be
	treated as a tuning parameter, we choose a Gaussian measure proportional to
	the Gaussian kernel $k_\theta(\cdot, \x^\star)$ to facilitate a similar closed-form
	solution:
	\begin{align} \label{eq:wimse}
		\text{wIMSE}&_n^{(m+1)} (\bar{\x}_{m+1}, \x^\star) 
		\equiv 
		\text{wIMSE}(\bar{\x}_{m+1},\X_n, \Y_n, \mathcal{X},\bar{\X}_m,\x^\star)
		 \\
		& \quad =\int_{\tilde{\x}\in \mathcal{X}}k_\theta(\tilde{\x},\x^\star)\frac{\sigma_{m+1,n}^2(\tilde{\x})}{\nu} \; d\tilde{\x} \nonumber \\
		& \quad =\frac{\sqrt{\theta\pi}}{2}\prod_{k=1}^d\Bigg(\text{erf}\left\{\frac{\x^\star-a_k}{\sqrt{\theta}}\right\}-\text{erf}\left\{\frac{\x^\star-b_k}{\sqrt{\theta}}\right\}\Bigg)-\text{tr}\Big\{\Big(\K^{-1}_{m+1}-\Q^{-1(n)}_{m+1}\Big)\W^*_{m+1}\Big\}, \nonumber 
	\end{align}
	where $\text{erf}$ is the error Gaussian function and $\W^\star_{m+1}=\prod_{k=1}^d \W^\star_{m+1,k}$. The $(i,j)^\text{th}$ entry of $\W^\star_{m+1,k}$ is
	\begin{align} \label{eq:wij} 
		w^{\star(i,j)}_{m+1,k}&\equiv w_{m+1,k}(\bar{\x}_{i}, \bar{\x}_{j}) \\
		&\;\; =\int_{a_k}^{b_k}k_\theta(\tilde{\x}_{k},\x^\star_k)k_\theta(\tilde{\x}_{k},\bar{\x}_{i,k})k_\theta(\tilde{\x}_{k},\bar{\x}_{j,k})\; d\tilde{\x}_{k} \nonumber \\
		&\;\; =\sqrt{\frac{\pi\theta}{12}}\text{exp}\Big\{\frac{2}{3\theta}\Big(\bar{\x}_{i,k}\x^*_k + \bar{\x}_{j,k}\x^*_k  + \bar{\x}_{i,k}\bar{\x}_{j,k}  - \x^{*2}_k - \bar{\x}_{i,k}^2 - \bar{\x}_{j,k}^2\Big)\Big\} \nonumber\\
		&\qquad \times\Bigg(\text{erf}\left\{\frac{\iota^{(u,j)}_{k}-3a_k}{\sqrt{3\theta}}\right\} -\text{erf}\left\{\frac{\iota^{(u,j)}_{k}-3b_k}{\sqrt{3\theta}}\right\}\Bigg), \nonumber
	\end{align}
	notating $\x^\star_k$ as the $k^\text{th}$ entry of the vector $\x^\star$ and $\iota^{(u,j)}_{k}=\x^\star_{k}+\bar{\x}_{u,k}+\bar{\x}_{j,k}$.
	Derivations for (\ref{eq:wimse}--\ref{eq:wij}) are included in Appendix
	\ref{app:wimse}. Extensions to other kernel structures, such as Mat\'ern
	\citep{Stein2012}, yield similar closed forms \citep[i.e., further
	extending][]{hetGP2}.
	
	The best new local inducing point can be found by solving the following program:
	\[
	\bar{\x}_{m+1} = \mathrm{argmin}_{\bar{\x}_{m+1} \in \mathcal{X}} \text{wIMSE}_n^{(m+1)} (\bar{\x}_{m+1}, \x^\star).
	\]
	The $\text{wIMSE}_n^{(m+1)} (\bar{\x}_{m+1}, \x^\star)$ surface realized over
	choices $\bar{\x}_{m+1} \in \mathcal{X}$, which we shall visualize momentarily
	in Section \ref{sec:illust}, may be multi-modal. However, it is not pathologically so
	like a global IMSE.  Library-based numerical schemes (details in Section
	\ref{sec:implement}) work well when suitably initialized but perform even
	better when aided by derivative information.  The $k^\mathrm{th}$ component of
	the gradient is given by
	\begin{align} \label{eq:dwimse}
		 \frac{\partial}{\partial\bar{\x}_{m+1,k}}
		&\text{wIMSE} (\bar{\x}_{m+1}, \x^\star)\\
		& = - \text{tr}\left\{\left(\frac{\partial \K^{-1}_{m+1}}{\partial\bar{\x}_{m+1,k}}-\frac{\partial\Q^{-1(n)}_{m+1}}{\partial \bar{\x}_{m+1,k}}\right)\W^\star_{m+1}\right\} - \text{tr}\left\{\left(\K^{-1}_{m+1}-\Q^{-1(n)}_{m+1}\right)
		\frac{\partial \W^\star_{m+1}}{\partial \bar{\x}_{m+1,k}}\right\}. \nonumber
	\end{align}
	The form of $\W^\star_{m+1}$, given in Eq.~(\ref{eq:wij}), reveals that
	the only non-zero entries in $\frac{\partial \W^\star_{m+1}}{\partial
		\bar{\x}_{m+1,k}}$ are the $m+1^{\text{st}}$ row/column. Those entries are
	\[
	 \frac{\partial w^{\star}_{m+1}(\bar{\x}_i,\bar{\x}_{m+1})}{\partial \bar{\x}_{m+1,k}}\prod_{k=1,k\neq k'}^d w^\star_{m+1,k}(\bar{\x}_i,\bar{\x}_{m+1}).
	\]
	Derivation of $\frac{\partial
		w^{\star}_{m+1}(\bar{\x}_i,\bar{\x}_{m+1})}{\partial \bar{\x}_{m+1,k'}}$ based
	on a squared exponential kernel is in Appendix \ref{app:wimse_grad}.
	
	Expressions for wIMSE and derivative (\ref{eq:wimse}--\ref{eq:dwimse})
	leverage the same Woodbury identities used earlier
	(\ref{eq:Sigi}--\ref{eq:GPpred}).  Partitioned inverse updates of
	$\K_{m+1}^{-1}$ and $\Q_{m+1}^{-1(n)}$ (Appendix \ref{app:KQpartitions}),
	allows $m\rightarrow m+1$ in $\mathcal{O}(m^2n)$ flops.
	
	\subsection{Illustrations of Greedy Inducing Point Search}
	\label{sec:illust}
	
\begin{figure*}[ht!]
		\begin{subfigure}{0.48\textwidth}
			\includegraphics[trim=15 15 20 48,clip, width=\textwidth]{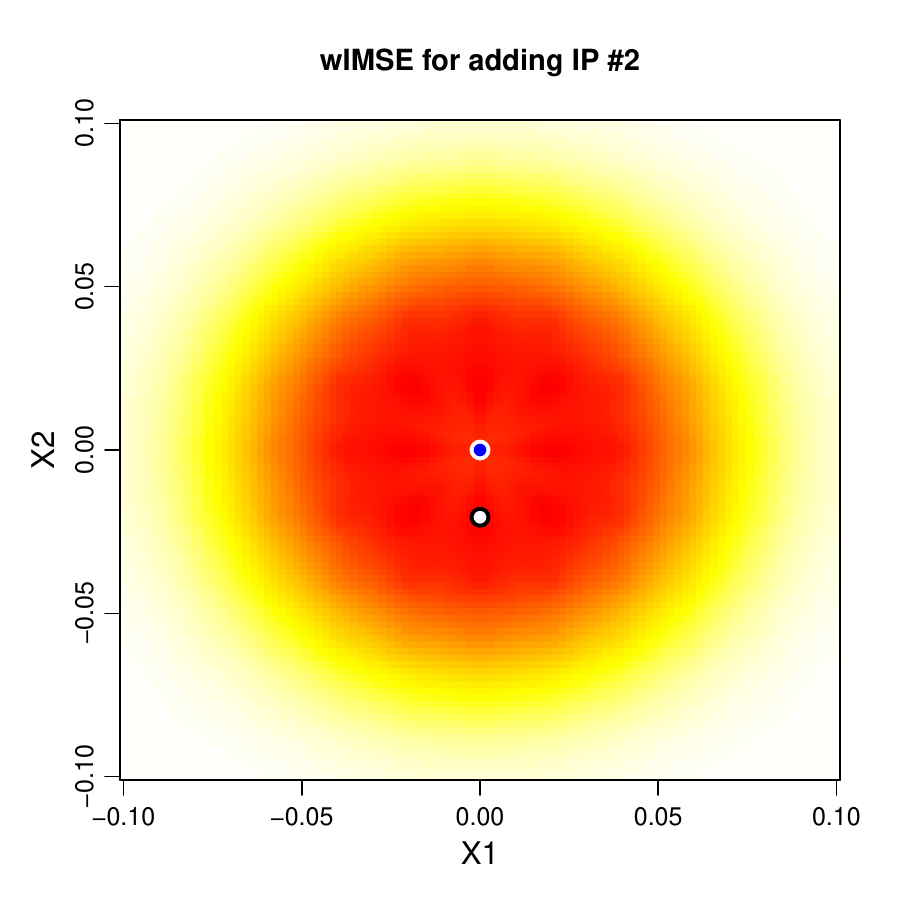}
			\caption{wIMSE for 2nd inducing point}
			\label{fig:4a}
		\end{subfigure}
		\hspace*{\fill}
		\begin{subfigure}{0.48\linewidth}
			\includegraphics[trim=15 15 20 48, clip,width=\textwidth]{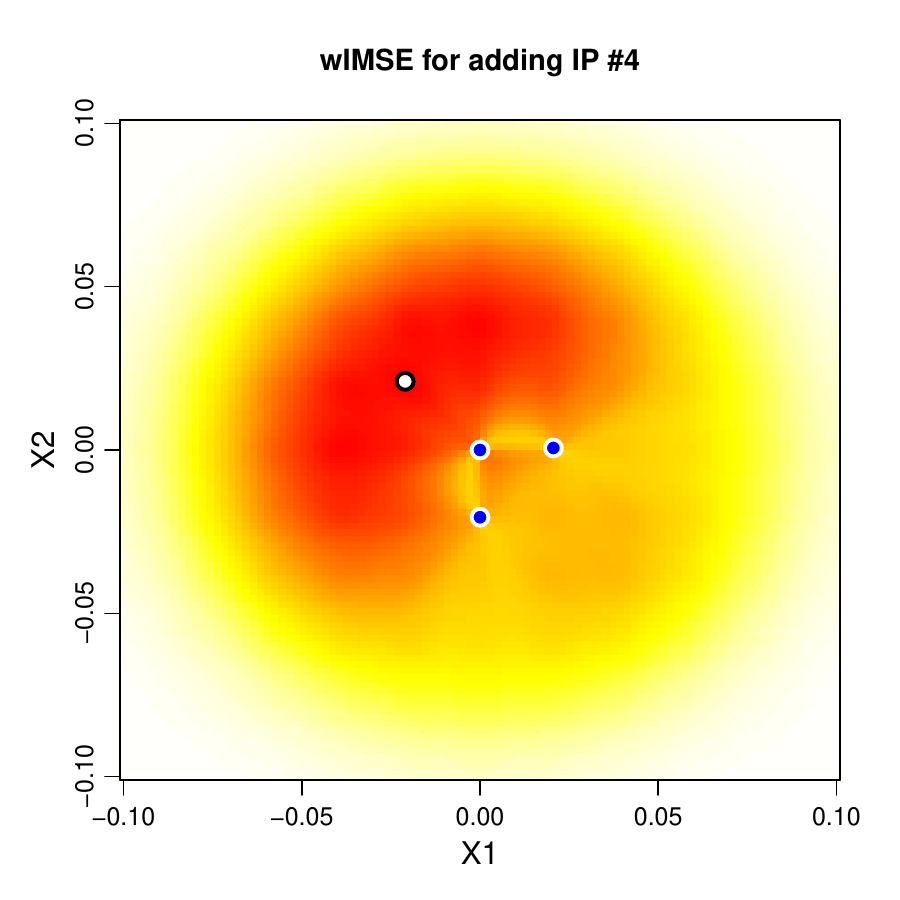}
			\caption{wIMSE for 4th inducing point}
			\label{fig:4b}
		\end{subfigure}\quad
		
		\begin{subfigure}{0.48\linewidth}
			\centering
			\includegraphics[trim=15 15 20 48, clip,width=\textwidth]{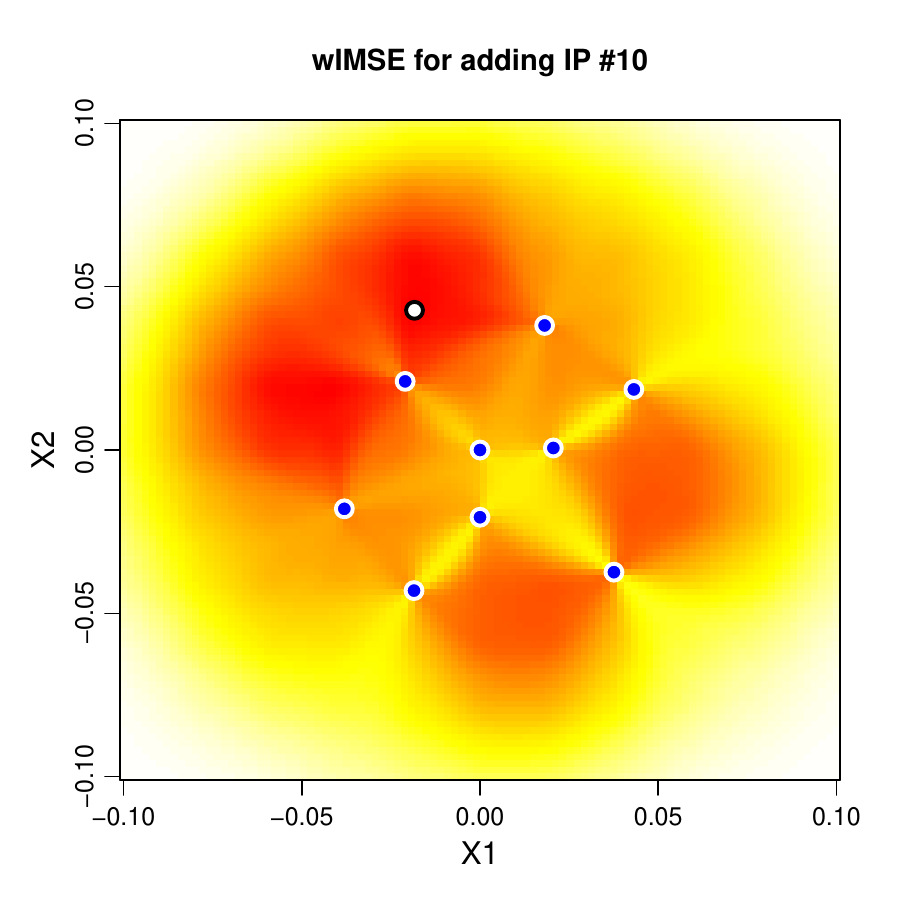}
			\caption{wIMSE for 10th inducing point}
			\label{fig:4c}
		\end{subfigure}\hspace*{\fill}
		\begin{subfigure}{0.48\linewidth}
			\centering
			\includegraphics[trim=15 15 20 48, clip,width=\textwidth]{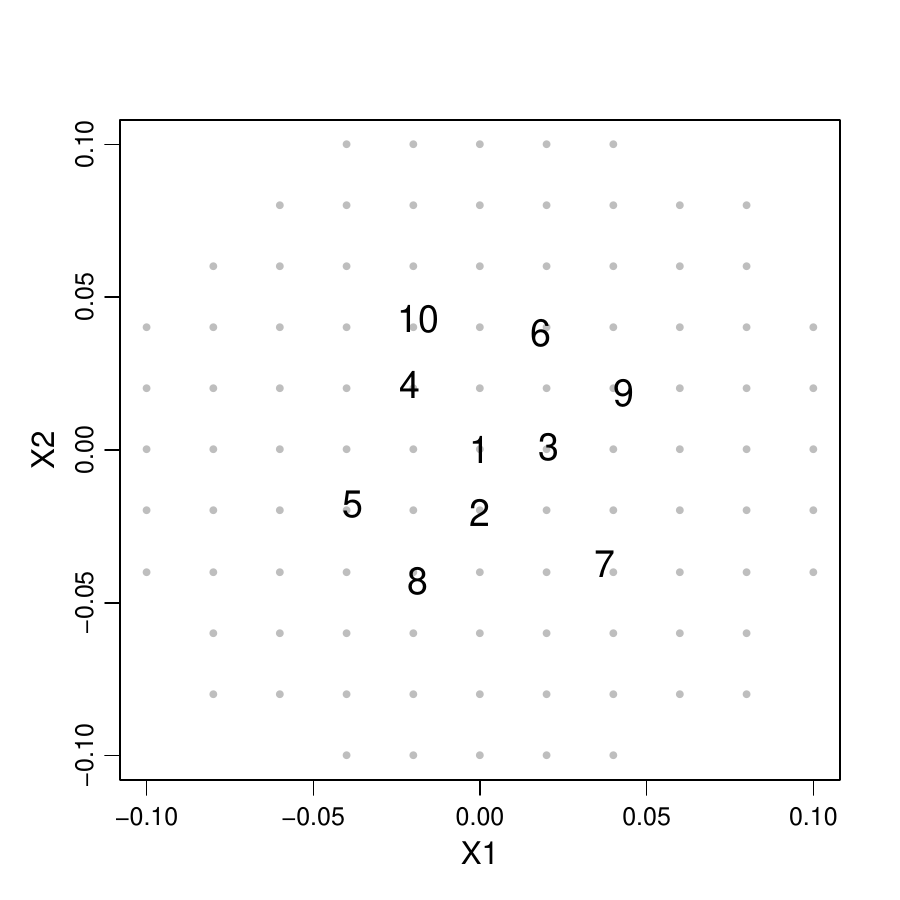}
			\caption{Local neighborhood and inducing points}
			\label{fig:center_ip_design}
		\end{subfigure}
		\caption{wIMSE surfaces (a--c), red/lower yellow/higher, used to optimize the
			2nd, 4th, and 10th inducing points: existing in blue; new selection in white.
			Predictive location $\x^\star$ is at the origin, which is where $\bar{\x}_1$
			is placed. Panel (d) summarizes the neighborhood $\X_n(\x^\star)$
			as gray dots and local inducing points $\bar{\X}_m(\x^\star)$ in number
			order.}
		\label{fig:wimse_seq}
	\end{figure*}

	Greedily optimizing wIMSE to place local inducing points around neighborhood
	$\X_n(\x^\star)$ results in $\bar{\X}_m(\x^\star)$ with (approximately)
	minimal predictive variance nearby $\x^\star$, so naturally they concentrate
	in that locale. To explore inducing point optimization with wIMSE, we use a toy 2d test problem
	known as Herbie's tooth \citep{herbtooth}. This function is attractive due to
	its low dimensionality but complex nonstationary surface littered with local
	minima. The function is defined by $f(x_1, x_2)= -w(x_1)w(x_2)$ where
	$w(x)=\text{exp}\left\{-(x-1)^2\right\}+\text{exp}\left\{-0.8(x+1)^2\right\}-0.05\sin\left(8(x+0.1)\right)$
	and $x_1,x_2 \in [-2,2]$. Figure \ref{fig:wimse_seq} shows the evolution of wIMSE-based
	acquisition for $\x^\star$ placed at the origin for Herbie's tooth ($N= 40$K,
	$n=100$).  Panels (a--c) show existing $\bar{\X}_m(\x^\star)$ in blue
	overlayed on the wIMSE surface used to select $\bar{\x}_{m+1}$. Optimal
	$\bar{\x}_{m+1}$, i.e., the wIMSE global minimum, are represented by
	white-filled circles. Unlike global VFE likelihood, ALC, and IMSE surfaces
	(explored in Appendix, \ref{app:imse}, Figure \ref{fig:optbakeoff}), the local wIMSE surface does not
	appear to be as affected by placement of the training points $\X_N$, or local
	neighborhood $\X_n(\x^\star) \subset \X_N$, shown as dots in panel (d). Local
	minima still exist as more inducing points are introduced. Yet the wIMSE
	surface is much smoother and well-behaved, making optimization easier.
	
	The first selection, $\bar{\x}_1(\x^\star)$, often lies very close to
	$\x^\star$. When $\x^\star$ is near the boundary of the input space, where
	wIMSE would be asymmetric, the first inducing point selection may ``pull
	away'' somewhat from $\x^\star$ towards to middle of the space.  But when
	symmetry is high, as it is at the origin for the illustration in Figure
	\ref{fig:wimse_seq}, it is hard to distinguish between $\bar{\x}_1$ and
	$\x^\star$ up to numerical error.  We find it convenient to simply begin
	optimizing at iteration two, with $\bar{\x}_1 = \x^\star$.
	
	\begin{algorithm*}[ht!]
		\caption{Inducing Point wIMSE Design}
		\label{alg:ip_wimse}
		\begin{algorithmic}[1]
			\Procedure {IP.wIMSE}{$m$, $n$, $\x^\star$, $\X$, $\mathcal{X}$}
			\State $\X_n \leftarrow$ NN$(\x^\star,\X, n)$ \hfill \Comment{Find $n$ nearest neighbors to $\x^\star$} 
			\State $\theta^{(0)} \leftarrow \mathrm{quantile}(0.1, \mathrm{dist}(\X_n))$\hfill
			\Comment{Reasonable local lengthscale} 
			\State $\bar{\x}_1 \leftarrow \x^\star$;\hfill \Comment{Place first inducing point}
			\For{$i=2,\dots,m$}\hfill \Comment{Greedy wIMSE to find the rest}
			\State  $\bar{\x}_i\leftarrow \mathrm{argmin}_{\bar{\x}_{i} \in \mathcal{X}} \text{wIMSE}_n^{(i)} (\bar{\x}_{i},\x^\star)$\hfill \Comment{Implicit dependence on $\theta^{(0)}$}
			\EndFor  \hfill \Comment{Implicit updates of local induced GP}
			\State \Return $\bar{\X}_{m}(\x^\star) = \{ \bar{\x}_i \}_{i=1}^m$ and $\X_n(\x^\star) = \X_n$
			\EndProcedure 
		\end{algorithmic}
	\end{algorithm*}
	
	For concreteness, steps for this greedy wIMSE inducing point search are
	outlined in Algorithm \ref{alg:ip_wimse}. After building the local
	neighborhood $\X_n(\x^\star)$, initialization is completed by choosing
	$\bar{\x}_1 \leftarrow \x^\star$ and local lengthscale $\theta^{(0)}$.
	Here we set $\theta^{(0)}$ based on quantiles of squared distances in
	$\X_n(\x^\star)$, though other settings are considered later. After greedy
	selection over $i=1,\dots, m$, intermixed with updates to the locally
	induced GP structure as outlined in Section \ref{sec:seqIP}, the procedure
	returns an $m\times d$ matrix comprised of the selected inducing points
	$\bar{\X}_m(\x^\star)$ alongside an $n \times d$ matrix defining the local
	neighborhood $\bar{\X}_n(\x^\star)$.
	
	\begin{figure*}[ht!]
		\centering
		\includegraphics[trim = 0 10 25 45, clip, width=0.48\textwidth]{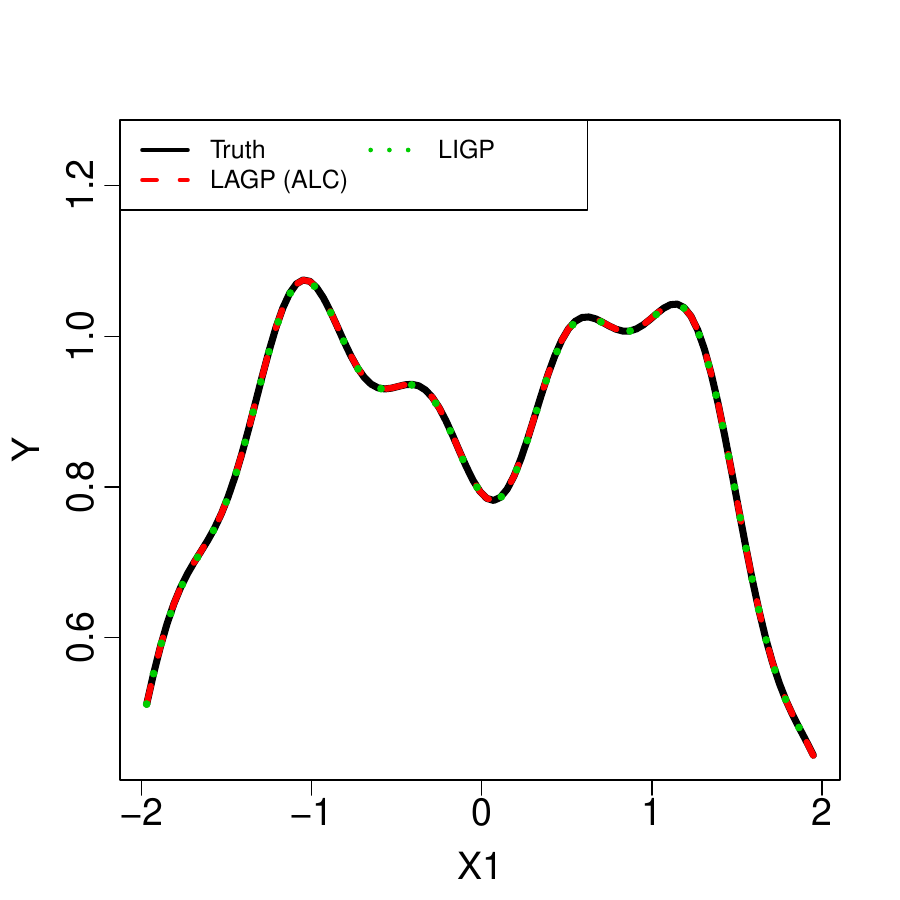} 
		\hspace{0.25cm}
		\includegraphics[trim = 0 10 25 45, clip, width=0.48\textwidth]{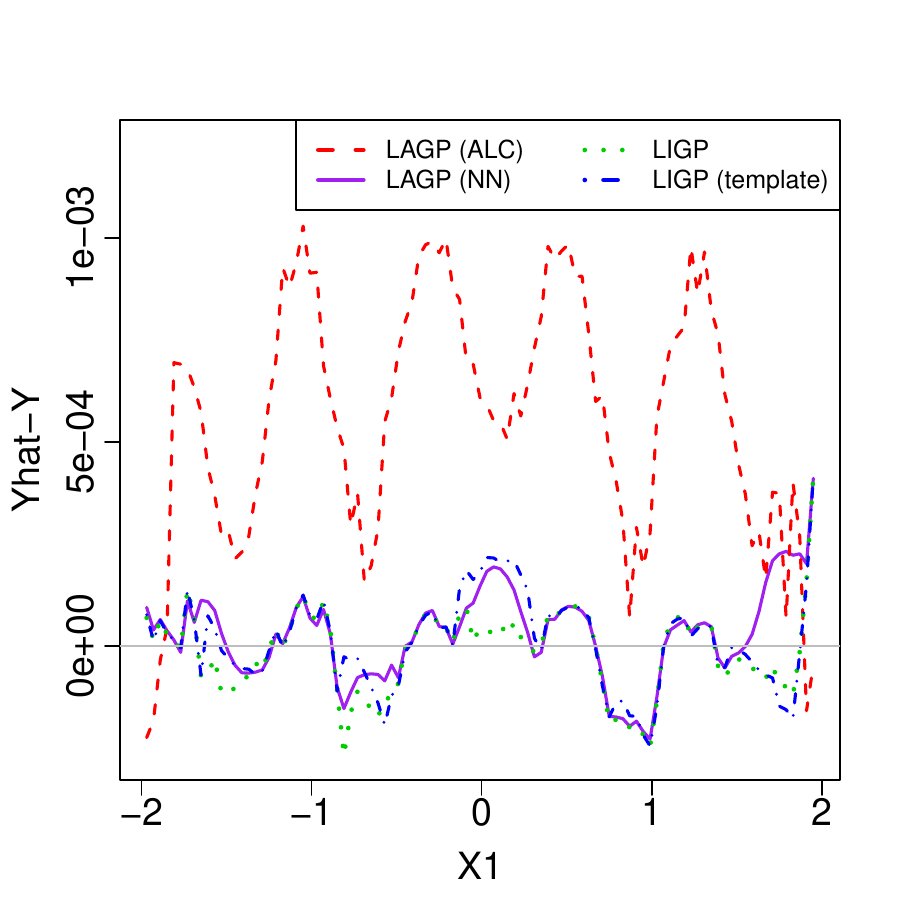}
		\caption{{\em Left}: approximate GP fits' mean prediction and truth on a slice of Herbie's
			tooth at $x_2^\star=0.6$. {\em Right}: errors relative to the truth on the approximate GP fits for the same slice of Herbie's tooth.}
		\label{fig:error_slice}
	\end{figure*}
	
	The left panel of Figure \ref{fig:error_slice} shows the predictions for a
	grid of $\x^\star$ settings arranged over a 1d slice of Herbie's tooth
	where $x_2^\star = 0.6$, including LAGP (via ALC with $n=50$, defaults in
	{\tt laGP}) and LIGP $(m,n)=(10,100)$, with local subset and inducing
	point designs re-optimized at each predictive location.  We allow LIGP a
	bigger neighborhood ($n$), with explanation in Appendix \ref{app:nsize}, but remind that this involves thriftier $m$-sized cubic
	decompositions.  Observe that both LAGP (red-dashed) and LIGP
	(green-dotted) capture the bumpiness of the surface, completely overlaying the true out-of-sample response (black-solid).
	
	Zooming in, the right panel of Figure \ref{fig:error_slice} shows errors along
	the slice under these comparators and two new variations: LAGP via NN with
	$n=100$ and LIGP with $(m,n)=(10,100)$ via template (Section
	\ref{sec:template}). Along most of the slice,  LIGP's error follows a similar
	trend as LAGP (NN, $n=100$), albeit with a bumpier line. This is not surprising
	given that both GP fits use the same neighborhood $\X_n(\x^\star)$. LAGP (ALC)
	copes well with smaller $n=50$ by filling $\X_n(\x^\star)$ with a mix of NNs
	and satellites.\footnote{ For identical $n$, ALC bests NN \citep{Gramacy2015},
		motivating increased $n$ for NN here.} Averaging along that slice,
	out-of-sample RMSE for LAGP (ALC) was $7.88\times
	10^{-4}$, versus $1.14\times10^{-4}$ and $1.12\times 10^{-4}$ for LAGP (NN)
	and LIGP, respectively. Here, LIGP predicts slightly better than LAGP (NN), its most direct competitor, and noticeably better than LAGP (ALC). By reducing the computational burden of the
	optimization criteria (NN v.~ALC) and matrix inversions (LIGP v.~LAGP), we free
	up resources to increase $n$ and thus accuracy.
	
	Encouraging as these early LIGP results are, selecting novel
	$\bar{\X}_m(\x^\star)$ for each $\x^\star$ is a substantial undertaking. LIGP
	required 3.32 seconds, on average, to greedily build $\bar{\X}_m(\x^\star)$
	using about 9 derivative-based iterates at each $\x^\star$.  Once in
	hand, optimizing via likelihood using a local analog of Eq.~\eqref{eq:SPGP}
	and predicting
	\eqref{eq:GPpred} based on $\bar{\X}_m(\x^\star)$ and $\X_n(\x^\star)$ is
	almost instantaneous, requiring 0.0062 seconds per prediction. LAGP (NN or
	ALC), which search discretely over subsets, lag a little behind at 0.0437 and 0.073 seconds, respectively.
	
	\section{Refinements to neighborhood composition}
	\label{sec:refine}
	
	LIGP can be accelerated with little impact on predictive accuracy
	by applying a single inducing point design $\bar{\X}_m(\x^\star)$ almost
	identically
	over all predictive
	locations $\x^\star \in \mathcal{X}$ of interest. Here we explore the benefits of inducing point design templates built with wIMSE and thriftier space-filling strategies.
	
	\subsection{Inducing points template}
	\label{sec:template} 
	
	Creating $\bar{\X}_m(\x^\star)$ based on wIMSE
	for each $\x^\star \in \mathcal{X}$ is a chore that can cannibalize any
	benefit that might come with adopting an inducing point approximation in the
	first place.  The highly structured nature of optimal wIMSE-based inducing
	points (Figure \ref{fig:wimse_seq}d) suggests such effort might be
	overkill.   Perhaps the cost of a single, representative optimization could be
	amortized over the expense of its application on a vast predictive grid.
	When re-purposed, through shifting or other transformation for new $\x^\star$,
	we refer to the original wIMSE design -- which might be calculated at the
	middle of the input space -- as a {\em template}.
	
	\begin{figure}[ht!]
		\centering
		\includegraphics[trim = 5 5 20 50, clip, width=.75\textwidth]{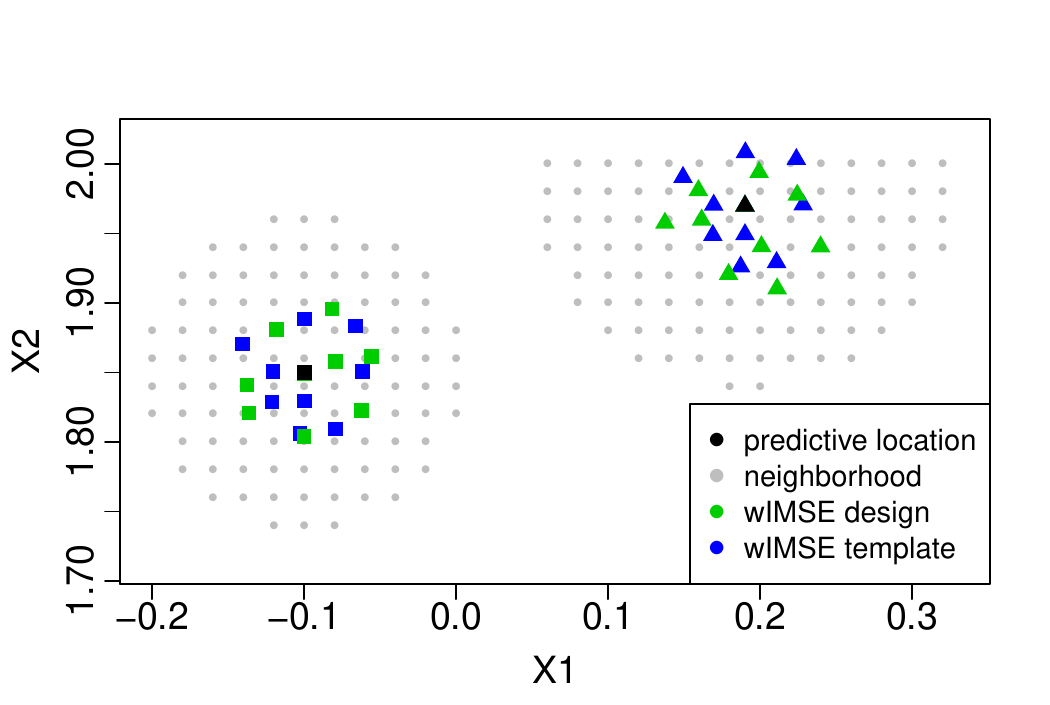}
		\caption{Local neighborhoods for two predictive locations $\x^{\star}$ at
			($-$0.1, 1.85) and (0.19, 1.97). Gray dots are $n=100$ neighborhoods
			$\X_n(\x^\star)$. Green points are wIMSE optimal inducing points
			$\bar{\X}_m(\x^\star)$; blue ones are displaced templates
			derived at the origin. The wIMSE template performs nearly the same space-filling effect as the locally optimized inducing points.}
		\label{fig:temp_compar}
	\end{figure}
	
	Figure \ref{fig:temp_compar} depicts the essence of the idea, comparing
	bespoke $\bar{\X}_m(\x^\star)$ to re-shifted ones from a template in two
	variations.  The setup is again Herbie's tooth in $[-2,2]^2$ and the two
	predictive sites are $\x^{\star(1)} = (-0.1, 1.85)$ and $\x^{\star(2)}=
	(0.19,1.97)$ whose $n=100$ neighborhoods $\X_n(\x^{\star(1)})$ and
	$\X_n(\x^{\star(2)})$, shown as gray dots, reside completely in the interior
	and on the $x_2$ boundary, respectively. Blue points in the plot represent a
	wIMSE-based inducing point design -- as optimized (Section \ref{sec:liGP}) at
	the center of the design space and then -- shifted to be centered at the
	$\x^\star$s. Compare these template-based local inducing points to
	corresponding optimal analogues in green. At both predictive locations, the
	pair of inducing point designs differ, yet both still space-fill the
	inner-neighborhood around $\x^\star$. A mild exception may be template-based
	$\bar{\X}_m(\x^{\star(2)})$ with its two points outside of the design
	region, which would not happen under an exhaustive re-optimization.  Other
	differences between alternatives would otherwise appear to be cosmetic up to
	rotation/small perturbations as may stem from a myriad of benign causes:
	relationship of $\x^\star$ to its local neighborhood $\X_n(\x^\star)$,
	convergence and global scope in greedy optimization, etc.
	
	Looking back at the right panel of Figure \ref{fig:error_slice}, observe how
	prediction errors based on templates (blue dashed line) compare with locally
	wIMSE-optimized inducing points (green dotted line) along the slice. Both LIGP
	variations seem to underestimate the response compared to LAGP (NN), but the
	template methods give nearly as accurate predictions as LIGP with locally
	wIMSE-optimized inducing points. Transferring a template captures most
	of the variability between local wIMSE designs, even at the boundaries. The
	template is also much faster. It took a total of 328.82 seconds to fit separate
	$\bar{\X}_m(\x^\star)$ and predict at the 99 $\x^\star$ locations depicted in
	the slice. Using a template instead takes 3.82
	seconds, a near two orders of magnitude improvement.
	
	\begin{algorithm}
		\caption{Building and Displacing Inducing Point Templates}
		\label{alg:ligp_temp}
		\begin{algorithmic}[1]
			\State $\check{\x} \leftarrow$ median($\X$) \hfill \Comment{Set $\check{\x}$ to the center of the data}
			\State $\bar{\X}_m \leftarrow$ {\sc IP.wIMSE}$(m,n,\check{\x},\X,\mathcal{X})$ \hfill \Comment{Use Alg. \ref{alg:ip_wimse} on $\check{\x}$}
			\State $\bar{\X}^{'}_{m} \leftarrow \bar{\X}_m - \check{\x}$ \hfill \Comment{Center template at the origin}
			\Procedure {IP.Template}{$n$, $\x^\star$, $\X$, $\bar{\X}_m'$}
			\State $\X_n \leftarrow \mathrm{NN}(x^\star, \X, n)$
			\State $\bar{\X}_m \leftarrow \bar{\X}^{'}_{m} + \x^\star$ \hfill \Comment{Simple displacement}
			\State \Return $\bar{\X}_{m}(\x^\star) = \{ \bar{\x}_i \}_{i=1}^m$ and $\X_n(\x^\star) = \X_n$
			\EndProcedure 
		\end{algorithmic}
	\end{algorithm}
	
	Algorithm \ref{alg:ligp_temp} provides pseudo-code for this template scheme,
	clarifying how a single wIMSE-based local inducing point design $\bar{\X}_m$
	is displaced for each $\x^\star$.  It is worth
	remarking that the scheme makes a tacit presumption that the full design
	structure, $\X_N$, is somewhat homogeneous: similar near the middle of the
	input space, $\check{\x}$, as near where it will be applied, i.e., for many
	disparate $\x^\star \in \mathcal{X}$.  We do not doubt it would be possible to
	engineer test problems, and/or non-space-filling designs $\X_N$, that
	would thwart this scheme, yet we find it works well in most cases.
	
	\subsection{Space-filling templates}
	\label{sec:spacefill}
	
	Our template-scheme leverages the neighborhood-focused space-filling nature of
	inducing points, beyond say $\bar{\x}_1 \approx \x^\star$. Space-fillingness is a
	cornerstone of (global) computer experiment design.  Numerous schemes exist,
	such as Latin hypercube samples \citep[LHSs][]{Mckay:1979} or maximin designs
	\citep{Johnson:1990}, etc., and hybrids thereof \citep{Morris:1995}.  These
	work well and often require less computation than model-based alternatives such as
	IMSE. If such space-filling designs (SFDs) could be re-tooled to ``focus'' on
	particular parts of the input space -- say in the neighborhood of $\x^\star$ --
	we might be able to avoid an expensive greedy wIMSE optimization all
	together.  SFDs might be able to mimic the behavior of a wIMSE template
	scheme at almost no cost at all.
	
	SFDs are usually constructed in a unit hypercube.  Re-centering such a
	template to $\x^\star$ is trivial, but re-scaling so that it lies within
	$\X_n(\x^\star)$ and resembles $\bar{\X}_m(\x^\star)$ is more challenging. One
	way is to derive a second, local rectangle as a means of defining a linear
	mapping between scales.	A thrifty strategy is to use the bounds of the neighborhood $\X_n(\x^\star)$.  But
	the shape of $\X_n(\x^\star)$ is roughly spherical, being comprised of
	Euclidean distance-based NNs. Thus the rectangular SFD will cover regions outside of the hypersphere, potentially placing some inducing points outside the
	neighborhood.  In low input dimension, say $d \leq 2$, this is no big deal,
	because the circumscription is relatively tight.  But when $d=8$, say, circumscription
	is poor.  
	
	\begin{figure*}[ht!]
		\centering				
		\begin{subfigure}{0.5125\textwidth}
			\includegraphics[trim=5 5 27 40, clip, width=\textwidth]{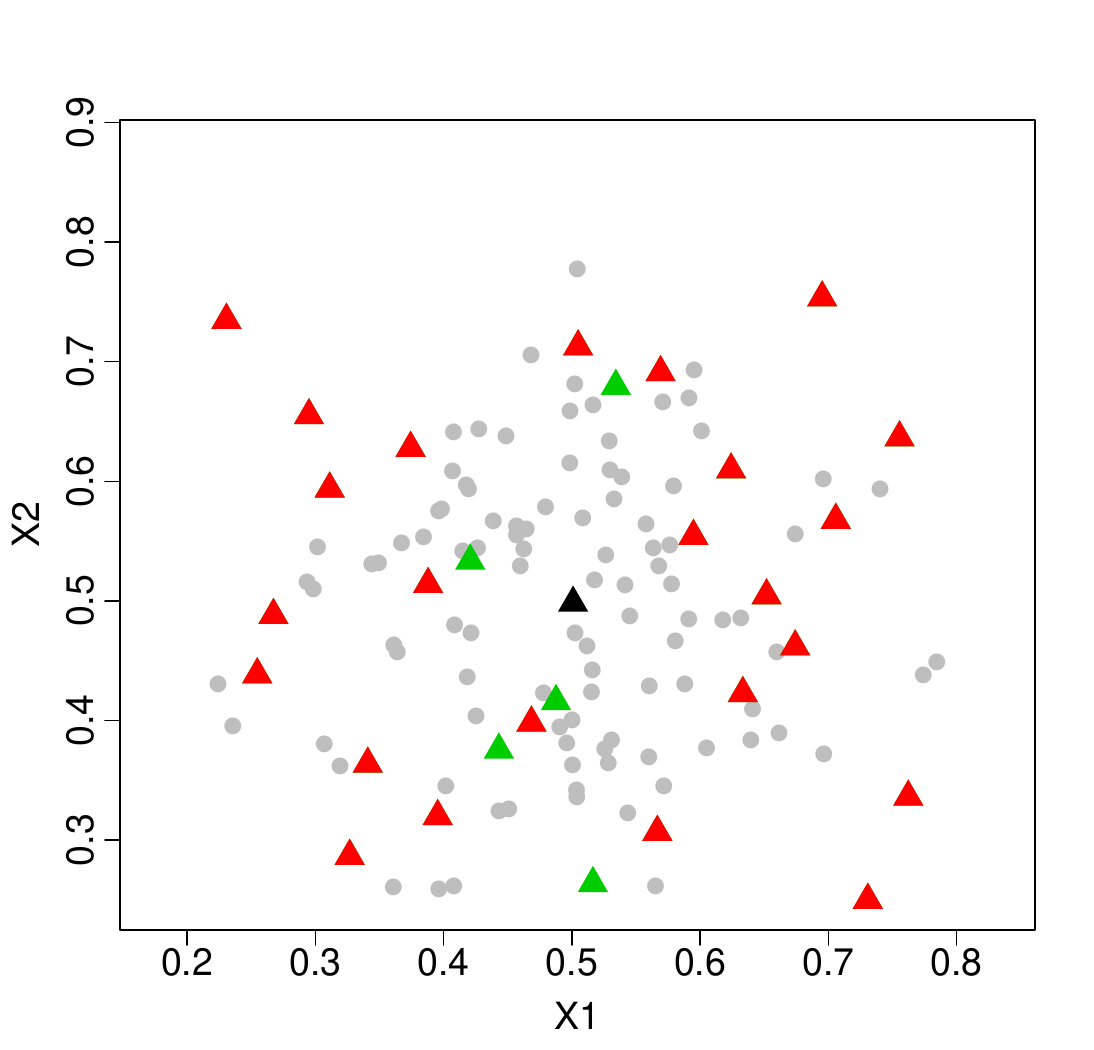}
			\caption{LHS rectangular template} \label{fig:lhs_temp}
		\end{subfigure}
		\begin{subfigure}{0.4675\textwidth}
			\includegraphics[trim=48 5 27 40, clip, width=\textwidth]{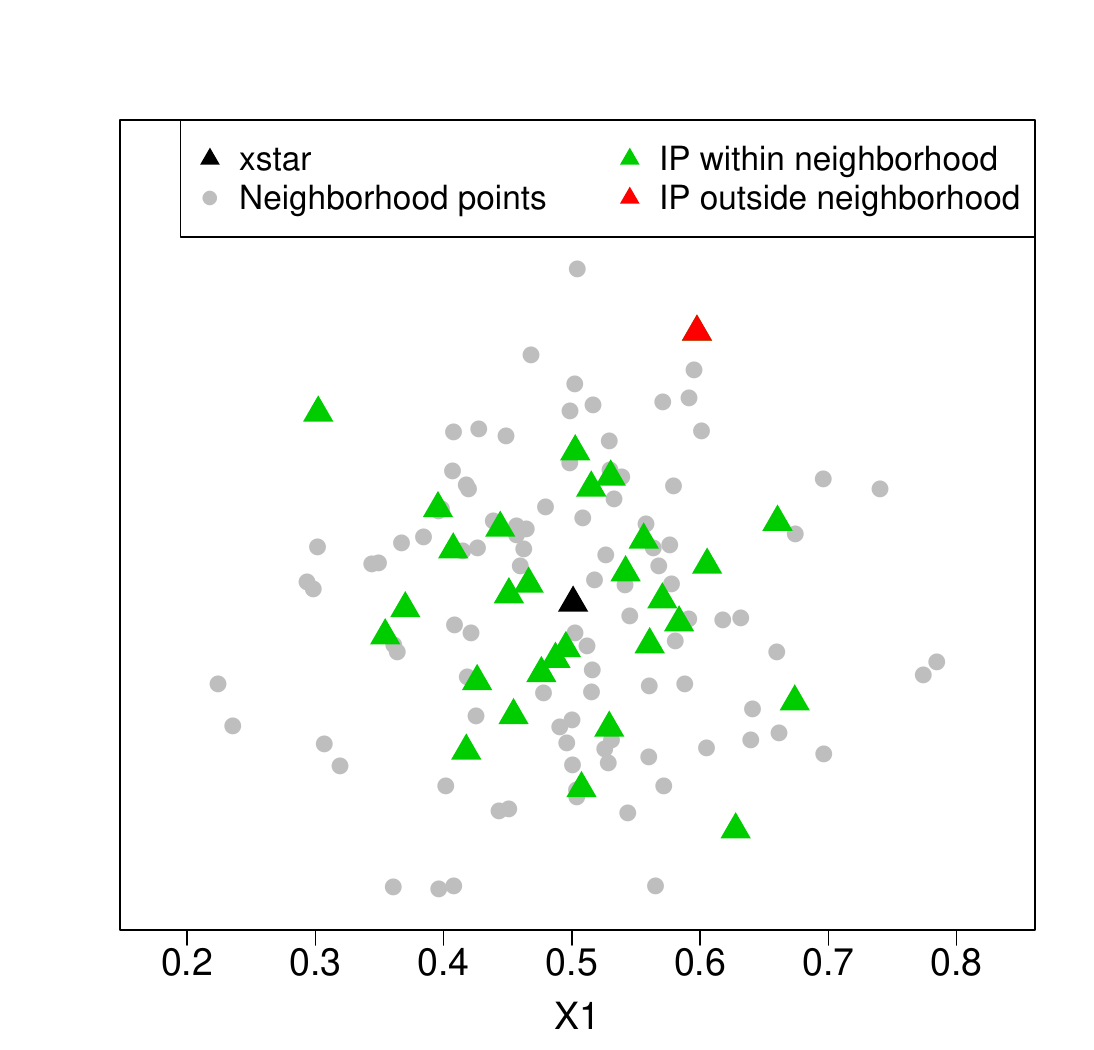}
			\caption{LHS $\Phi$ template} \label{fig:qnorm_temp}
		\end{subfigure}
		\hspace*{\fill}
		\caption{SFD template schemes (triangles) in 2d projections relative to local
			neighborhood (gray dots): (a) rectangular re-scaled LHS template (triangles)
			 in relation to a local neighborhood (gray dots); (b) qNorm LHS
			template. Green triangles indicate $\bar{\X}_m(\x^\star)$ within the
			neighborhood $\X_n(\x^{\star})$ in all coordinates; red outside.}
		\label{fig:space_fill_temp}
	\end{figure*}
	
	Figure \ref{fig:space_fill_temp}a shows a 2d projection of an 8d local
	neighborhood for the borehole problem, described in Section
	\ref{sec:borehole}.  Here, the volume of the convex hull of the neighborhood
	$\X_n(\x^\star)$ is less than one sixtieth of the size of the rectangle
	circumscribing its bounds in the coordinate axis directions. Consequently many of
	the template re-scaled local inducing points $\bar{\X}_m(\x^\star)$, indicated
	as triangles, lie outside the neighborhood (red) in at least one of the eight
	coordinates.  Of the $m=30$ local inducing points calculated for that figure,
	one of which is automatically at $\x^\star$, only five rectangular re-scaled
	LHS template points lie within the neighborhood.

	As remedy, we propose a nonlinear mapping that warps the SFD to lie inside the
	neighborhood with high probability.  In particular, we scale the SFD based on
	an inverse Gaussian CDF ($\Phi^{-1}$), applied separately to each of the $d$
	input coordinates. Algorithm \ref{alg:qnorm_design} outlines steps towards
	generating an inducing point design $\bar{\X}_m(\x^\star)$ based on a
	SFD $\hat{\X}$ of size $m-1$, i.e., beyond choosing
	$\bar{\x}_1 = \x^\star$. $\Phi^{-1}$ calculations for each dimension
	$k=1,\dots,d$ involve $\mu = \x^\star_k$ and variance $\theta^{(0)}$.  This is
	the same $\theta^{(0)}$ as in Algorithm \ref{alg:ip_wimse} for greedy wIMSE
	optimization, except here we demonstrate a more absolute default choice. This
	$\Phi^{-1}$ transformation yields higher density near $\x^\star$ and much
	lower density outside of the neighborhood's hypersphere. Observe in Figure
	\ref{fig:space_fill_temp}b how this warping drastically reduces the number
	of template points outside of the neighborhood.
	
	\begin{algorithm*}[ht!]
		\caption{Inverse Gaussian CDF Space-Filling Template}
		\label{alg:qnorm_design}
		\begin{algorithmic}[1]
			\Procedure {IP.qNorm}{$m$, $n$, $\x^\star$, $\X$}
			\State $\X_n \leftarrow$ NN$(\x^\star,\X, n)$ \hfill \Comment{Find $n$ nearest neighbors to $\x^\star$} 
			\State $\theta^{(0)} \leftarrow (\frac{1}{3} \max_{k} |\X_{n,k}(\x^\star) -  x^\star_k|)^2$ \hfill
			\Comment{Reasonable local lengthscale} 
			\State $\hat{\X} \leftarrow \text{SFD}[0,1]^d \text{ with } m-1 \text{ points}$ 
			\hfill \Comment{Could be moved outside} 
			\For{$k=1,\dots,d$}\hfill \Comment{Warp each input coordinate}
			\State  $\breve{\x}_d \leftarrow \Phi^{-1}(\hat{\x}_d;\mu=x^\star_d, \sigma^2=\theta^{(0)}) $ \hfill \Comment{Inverse Gaussian CDF with $\mu,\sigma^2$}
			\EndFor
			\State $\bar{\X}_{m} \leftarrow \text{rowbind}(\x^\star, \breve{\X})$ \hfill \Comment{Add $\x^\star$ as inducing point}
			\State \Return $\bar{\X}_{m}(\x^\star) = \bar{\X}_m$ and $\X_n(\x^\star) = \X_n$
			\EndProcedure 
		\end{algorithmic}
	\end{algorithm*}
	
	Pseudocode in Algorithm \ref{alg:qnorm_design} conveys bespoke SFD within each
	application of the subroutine, yielding new $\hat{\X}$ in each call.  As with
	the wIMSE template in Algorithm \ref{alg:ligp_temp}, this can be moved
	outside the subroutine to fix a single SFD, which might be important if the
	SFD is expensive to compute. We prefer LHSs for our SFDs because they are
	easy/instantaneous via libraries such as {\tt lhs} \citep{lhs} on CRAN.
	Hybrids such as maximin--LHS are also straightforward (also with {\tt lhs}),
	which can avoid some pathologies inherent in random LHS design. Ordinary
	maximin can be problematic under $\Phi^{-1}$ because that criteria places
	points on the bounding hypercube, which would warp to $\pm \infty$ without
	intervention, and because evaluating and optimizing that criteria is slow.
	Uniformly random design may be preferred when local lengthscales are difficult
	to estimate \citep{zhang2018distance}.
	
	Section \ref{sec:illust} offered comparison between run time and predictive
	accuracy for LIGP, using wIMSE to build unique inducing point designs, to that
	of LAGP on a slice of Herbie's tooth. Now consider new template comparators:
	hyperrectangular SFD, LIGP (cHR), and $\Phi^{-1}$-scaled SFD, LIGP (qNorm).
	While it took 3.32 seconds on average to build wIMSE-based designs, scaling an
	SFD to circumscribe the neighborhood (cHR) or applying $\Phi^{-1}$ (qNorm)
	only takes 0.01 seconds on average. Both of these SFD template schemes produce
	an RMSE that is essentially the same ($1.8\times10^{-4}$) as applying the
	wIMSE template scheme.
	
	The borehole problem uses larger $(m,n) = (80,150)$ settings due to the higher
	input dimension (see Appendix \ref{app:nsize} for a discussion).  It takes 141 seconds to build a wIMSE-based inducing
	point template of size $m$, while it only takes 0.034 seconds to build a
	SFD-scaled template. SFD and wIMSE templates produce LIGPs with
	similar RMSEs, discussed in Section \ref{sec:borehole}.

	\section{Computation and benchmarking}
	\label{sec:examples}
	
	Here we provide implementation details followed by in-depth comparison of
	LIGP and various template schemes, to LAGP on a swath of
	synthetic and real computer simulation experiments. Our metrics
	for benchmarking are out-of-sample RMSE and computation time. All analysis
	was performed on an eight-core hyperthreaded Intel i9-9900K CPU at 3.60GHz.
	
	\subsection{Implementation details}
	\label{sec:implement}
	
	{\sf R} code \citep{R} supporting our methodological contribution, and all
	examples, may be found on our Git
	repository.
	\begin{center}
		\url{https://bitbucket.org/gramacylab/lagp/src/master/R/inducing/}
	\end{center}
	Some noteworthy aspects of that implementation include the following.  Unlike
	{\tt laGP}, which is coded in {\sf C} with {\tt OpenMP} for symmetric
	multiprocessing parallelization ({\sf R} serving only as wrapper),
	our LIGP implementation is pure {\sf R}.  Nevertheless, our template
	schemes are competitive, time-wise, and sometimes notably
	faster.
	
	We privilege an isotropic Gaussian kernel formulation with scalar
	lengthscale $\theta$ for local modeling, although there is no reason other
	forms, such as Mat\'ern \citep{Stein2012}, could not be entertained so
	long as the structure is differentiable with respect to inducing points
	$\bar{\X}_m$.  To improve numerical conditioning of matrices $\K_m$ and
	$\Q_m^{(n)}$ for stable inversion, we augment their diagonals with
	$\epsilon_K=10^{-6}$ and $\epsilon_Q=10^{-5}$ jitter \citep{Neal1998},
	respectively. While both are theoretically decomposible, we find that
	$\Q_m^{(n)}$ is more sensitive to conditioning issues, thus requiring
	larger $\epsilon$. In the context of LAGP, it has been shown that
	separable local formulations do not much improve predictive performance,
	especially after first applying a global pre-scaling of inputs
	\citep{sun2019emulating}. Such stretching and compressing of
	inputs,\footnote{A characterization attributed to Derek Bingham predating
	any published account, to our knowledge.} has recently become popular as a
	means of boosting predictive performance of approximate GP methods
	\citep[e.g.,][]{katzfuss2020scaled}. When pre-scaling in our exercises to
	ensure apples-with-apples comparisons to benchmarks we divide by
	square-root separable global lengthscales obtained from a GP's fit to
	random size-1000 data subsets. See \cite{gramacy2020surrogates}, Section
	9.3.4, for details. The time required is not included in our summaries.
	
	Building of wIMSE inducing point designs $\bar{\X}_m(\x^\star)$ and templates
	$\bar{\X}_m(\check{\x})$, generically $\bar{\X}_m$ below, follows Algorithm
	\ref{alg:ligp_temp} with $m$ and $n$ appropriate to the input
	dimension $d$ (Appendix \ref{app:nsize}), provided momentarily with our
	particular exercises. For initial local lengthscale $\theta^{(0)}$, we have
	had success with a number of heuristics which often lead to similar
	values/performance for LIGP methods in our exercises.
	\cite{laGP} suggests the 10\% quantile of squared pairwise
	distances between the neighborhood points $\X_n$.\footnote{In {\tt laGP}, the
		function providing $\theta^{(0)}$ in this way is {\tt darg}.} See
	Algorithm \ref{alg:ip_wimse}. A downside is that this is quadratic in
	$n$.  A more absolute/direct $\mathcal{O}(n)$ approach matches $\theta^{(0)}
	= \sigma^2$, where $3\sigma$ approximates the 99\% quantile of a Gaussian fit,
	to the margins of $\X_n$. Algorithm \ref{alg:qnorm_design} exemplifies this
	choice for contrast, although we see these as interchangeable. Each
	$\bar{\x}_{m+1}$ augmenting $\bar{\X}_m$ optimizing wIMSE is found via
	a 20-point multi-start L-BFGS-B \citep{Byrd1995} scheme (using {\tt optim} in
	{\sf R}) peppered within the bounding box surrounding the neighborhood $\X_n$
	to a tolerance of 0.01.	Templates derived from space-filling  designs (Section \ref{sec:spacefill}) originate from $m-1$ point LHSs through the hyperrectangle enclosing
	$\X_n(\check{\x})$, and then augmented with $\check{\x}$ as the
	$m^\mathrm{th}$ inducing point.

	Regardless of inducing point/template construction, machinery behind
	LIGP-based prediction is identical.   Algorithm \ref{alg:ligp_pred} outlines
	the steps to construct local neighborhoods and predict at each of a set of
	$N'$ prediction locations $\X^\star$ given training data $\{\X_N,\Y_N\}$,
	neighborhood size $n$, and number of inducing points $m$. Each location
	$\x_i$, for $i = 1, \dots, N'$ could proceed in parallel.  In our
	implementation we use 16 threads.\footnote{I.e., two per hyperthreaded core.}
	The pseudocode attempts to be agnostic about the inducing point scheme by
	simply writing $\mathrm{IP}(\dots)$.  Any of Algorithms
	\ref{alg:ip_wimse}--\ref{alg:qnorm_design} can be used here. To estimate
	scale and lengthscale we used
	Eqs.~(\ref{eq:nuhat}--\ref{eq:concentrate_ll}) through simple substitutions of
	$(m,n)$ for the local neighborhoods of $\x^\star$. We rely on {\tt optim} in
	{\sf R} to minimize the negative log-likelihood to estimate local
	$\hat{\theta}(\x^\star)$'s. Finally, the predictive mean and variance for
	$\x^\star$ are extracted via Eq.~\eqref{eq:GPpred}.
	
	\subsection{Borehole}
	\label{sec:borehole}
	
\begin{figure*}[ht!]
		\centering
		\begin{subfigure}{0.49\textwidth}

			\includegraphics[trim=0 30 0 4, clip, width=\textwidth]{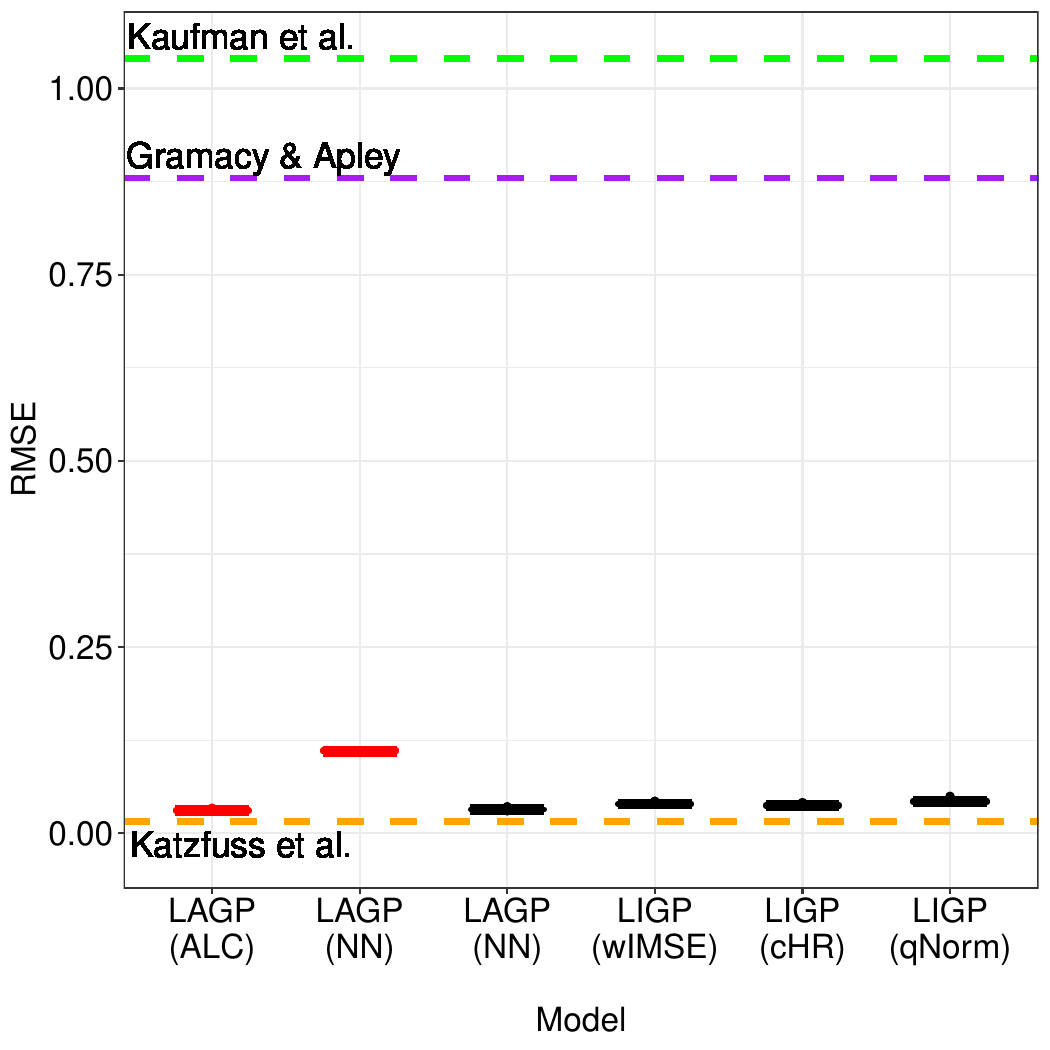}
		\end{subfigure}
		\begin{subfigure}{0.49\textwidth}
			\includegraphics[trim=0 30 0 4, clip, width=\textwidth]{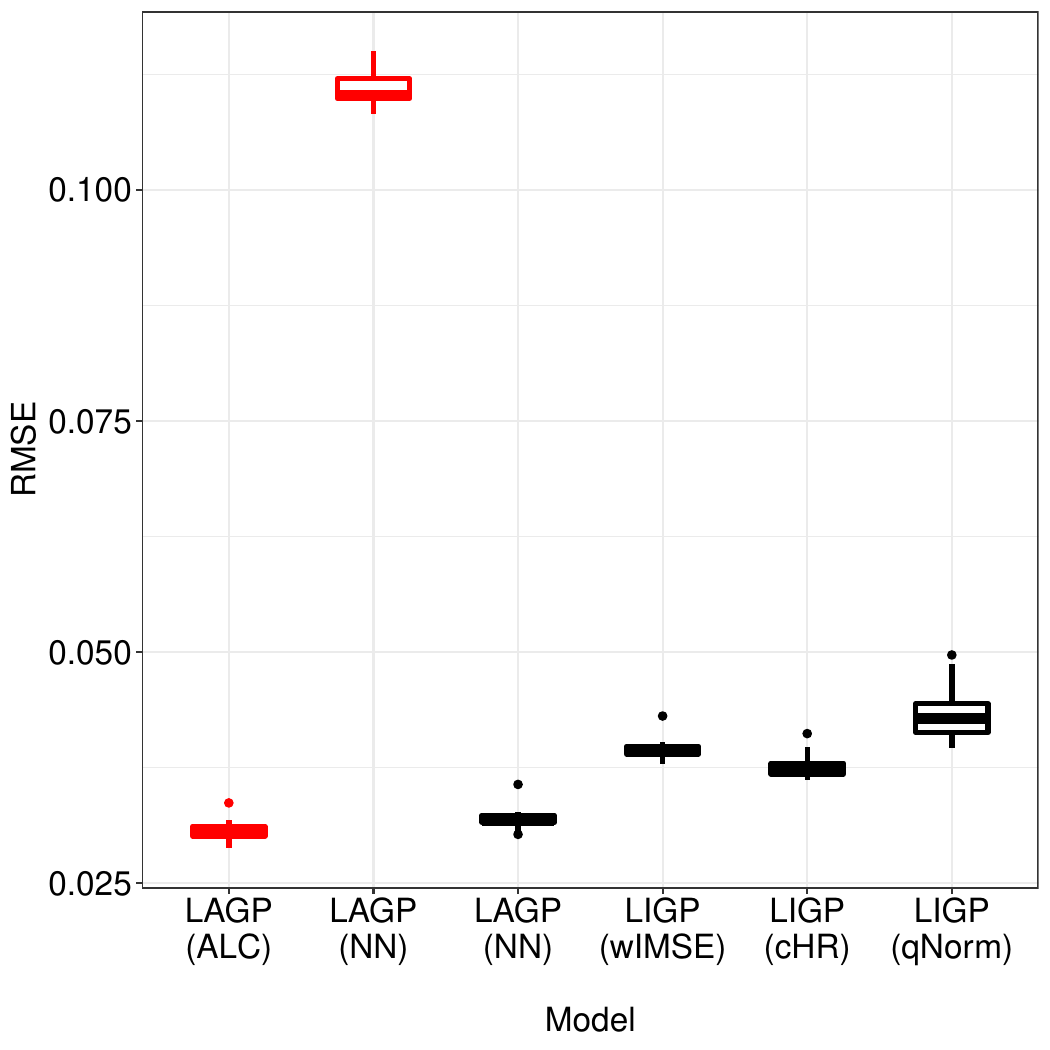} 	
		\end{subfigure}
		\vspace{0.25cm}
		\begin{tabular}{l|l|l|l|l|l|r} 
			\hline
			Method & LAGP & LAGP & LAGP & LIGP & LIGP & LIGP\\ 
			\hline
			Variation & ALC & NN & NN & wIMSE & cHR & qNorm \\ \hline
			$n$ & 50 & 50 & 150 & 150 & 150 & 150 \\ \hline
			Time & 0.97 & 0.08 & 1.27 & 3.06 & 0.73 & 0.73
		\end{tabular}
		\caption{{\em Top-left:} accuracy over 30 MC repetitions with lines showing
			other published works' results: \cite[][green]{kaufman2011efficient},
			\cite[][purple]{Gramacy2015}, \cite[][orange]{katzfuss2020scaled}. {\em
				Top-right:} zoomed in version focusing on the best LI/LAGP methods. The color
			of the boxplot outline, red and black, correspond to the sizes of the neighborhoods
			($n=50,150$ respectively). {\em Table below:} compute time in minutes.}
		\label{fig:borehole_boxplot}
	\end{figure*}

	Previewed in Section \ref{sec:spacefill}, the borehole function
	\citep{borehole} is a classic example in computer experiments literature.
	Outputs may be derived in closed form as
	$$y=\frac{2\pi
		T_u[H_u-H_l]}{\log\left(\frac{r}{r_w}\right)\Big[1+\frac{2LT_u}{\log(r/r_w)r_w^2K_w}+\frac{T_u}{T_l}\Big]}$$
	via inputs in the eight-dimensional rectangle: 
	\begin{align*}
		r_w&\in[0.05,0.15] & r&\in[100,5000] 
		&T_u&\in[63070,115600] &
		T_l&\in[63.1,116] \\ 
		H_u&\in[990,1100] & H_l&\in[700,820] 
		&L&\in[1120,1680] &
		K_w&\in[9855,12045]. 
	\end{align*} 
	For training we use LHSs of size $N = 100000$, recoding natural inputs to the
	unit 8-cube followed by pre-scaling via a global separable
	$\hat{\theta}$ as explained in Section \ref{sec:implement}. 
	We use $(m,n)=(80,150)$ for all LIGP fits (see Appendix \ref{app:nsize}). For a fair comparison, we entertain
	$n=150$ for LAGP (NN) as well as the default of $n=50$ for  NN and
	ALC-based LAGP comparators. Figure \ref{fig:borehole_boxplot} summarizes RMSEs obtained over thirty MC
	instances with novel training and $N'=10000$ sized LHS testing sets.
	
	Mirroring other studies \citep[e.g.,][]{sun2019emulating}, local approximation
	is key to using a vast training data set to get good predictions.  LAGP
	performs better with a neighborhood of $n=50$ selected using ALC versus even
	larger neighborhoods ($n=150$) using NN. Given the smoothness of the borehole
	surface, the addition of ``satellite" points provided by ALC gives an accuracy
	boost over pure NN of similar size. We believe the same to be true of
	LIGP (cHR). Any inducing points lying outside the neighborhood act as
	``satellites'' in this context. This is backed up by comparable RMSE results.
	The added flexibility of inducing points (LIGP) over discrete subsets (LAGP)
	may be limited by the highly smooth borehole dynamics.
	
	Timings are provided at the bottom of Figure \ref{fig:borehole_boxplot}, with
	LIGP LHS templates being fastest among the most competitive alternatives,
	accuracy-wise. Interestingly, the cHR template is even better at prediction than the
	optimized wIMSE one, obtained at great computational expense (3.06 minutes).
	Compared to LAGP (NN) with $n=150$, accuracy is only slightly diminished, but
	predictions are furnished in half the time on aggregate.  Again, we remind the
	reader that this is a little unfair to LIGP, comparing an {\sf R}-only
	implementation to {\tt laGP}'s {\sf C} library.  Another reason this timing
	comparison is not more impressive is that optimizing the inducing point
	likelihood to obtain local $\hat{\theta}(\x^\star)$, despite being cubic in
	$m$ rather than $n$, tends to take more BFGS iterations than the LAGP analog.
	
	Although LIGP methods do not best LAGP (except NN with $n=50$) on
	accuracy, it is important to place these RMSEs in context.  Horizontal
	dashed lines in the left panel of Figure \ref{fig:borehole_boxplot} offer
	wider historical perspective.  \cite{kaufman2011efficient}'s reported an
	RMSE of 1.4 (green line; 99\% sparse) with $(N,N')=(4000, 500)$ in 17
	minutes via compactly supported kernels. \citet{Gramacy2015}'s initial
	LAGP (ALC) implementation improved that to 0.88 (purple line) in 3
	minutes, utilizing eight cores.  Subsequent improvements in handling
	larger (less well-conditioned) matrices, and wider {\tt OpenMP}
	parallelization bring us to the orders of magnitude more accurate and fast
	results in Figure \ref{fig:borehole_boxplot}.
	
	More recently a method called SVecchia \citep{katzfuss2020scaled}, adapted
	from geostatistcs to computer surrogate modeling, has yielded
	impressive RMSEs of 0.016 (orange line) in similar exercises ($(N,
	N')=(100000,20000)$) in about five minutes -- combining training (4.4 minutes)
	and testing (0.4 minutes) phases -- in a single-core setting.  We see this new
	vanguard of methods as equivalent on the borehole problem, with
	nuance depending on the application.  For example, if you need a
	one-off prediction, LAGP methods (e.g., ALC) are best, furnishing accurate
	predictions in fractions of a second without an explicit training phase.  With
	modest testing sizes, LIGP methods are faster when amortizing the cost of
	template calculation.  For larger testing sets, SVecchia methods seem
	attractive.
	
	Lastly, consider comparing to a more traditional global form of inducing point
	prediction (Section \ref{sec:opt}). Using an LHS for $\bar{\X}_M$ with $M=80$ in
	$[0,1]^8$ requires only 0.56 minutes to produce predictions
	\eqref{eq:GPpred} with fixed lengthscale $\theta$, less than even the
	space-filling template variations of LIGP.  Accuracy is tightly coupled to
	$\theta$, but MLEs render the method uncompetitive as a single evaluation of the
	log-likelihood \eqref{eq:concentrate_ll} takes nine minutes.
	
	\subsection{Robot arm}
	\label{sec:sarcos}
	
	The SARCOS data is a popular computer simulation benchmark from the machine
	learning literature \citep{vijayakumar2000locally,Rasmussen2005}. The
	data/simulations\footnote{Original MATLAB:
		\url{http://www.gaussianprocess.org/gpml/data/}; plain text in our
		Git repo.} model seven torque outputs as a function of 21 input variables
	consisting of position, velocity, and acceleration of a robot arm.  It comes
	pre-partitioned into a training set of size $N=44484$ and a testing set of
	size $N'=4449$. Here we consider only the first torque output.  High input
	dimensionality and non-uniform design -- inputs lie on a low-dimensional
	manifold in the input space -- present surrogates with unique challenges.
	
	One implication of the non-uniform design for LIGP is that a
	hyperrectangle surrounding $\X_n(\check{\x})$, for median input $\check{\x}$,
	does not place $\check{\x}$ in its center. Consequently a cHR template
	would yield an un-centered $\X_m(\x^\star)$.
	Space-fillingness is preserved, albeit with many points outside of the
	hypersphere enclosing $\X_n(\check{\x})$. A qNorm template, by contrast,
	can preserve centering through $\Phi^{-1}$. However, in both cases the
	low-dimensional input manifold may result in a fair number of inducing points
	without many $\X_n(\x^\star)$ nearby.
	
	As with previous examples, we perform an input pre-scaling based on separable
	lengthscales estimated via MLE from a size $n=1000$ random data subset.  After
	pre-scaling we find that local likelihoods, for both LAGP and LIGP, are flat
	for many $\x^\star$, yielding exceedingly long local lengthscales
	$\hat{\theta}(\x^\star)$ and ``washed out'' local surrogates.  Apparently, in
	21 input dimensions, small neighborhoods ($n=50$ and $n=200$) provide
	insufficient information about local lengthscales, i.e., beyond the global one.
	Although we show results with LAGP in both variations, with and without local
	MLE calculations (with both isotropic and separable local kernels), all
	variations entertained perform much better with a fixed $\theta_0=1$ for all
	local calculations.  
	
	\begin{figure}[ht!]
		\centering
		\includegraphics[trim=3 5 10 50, clip, width=0.75\textwidth]{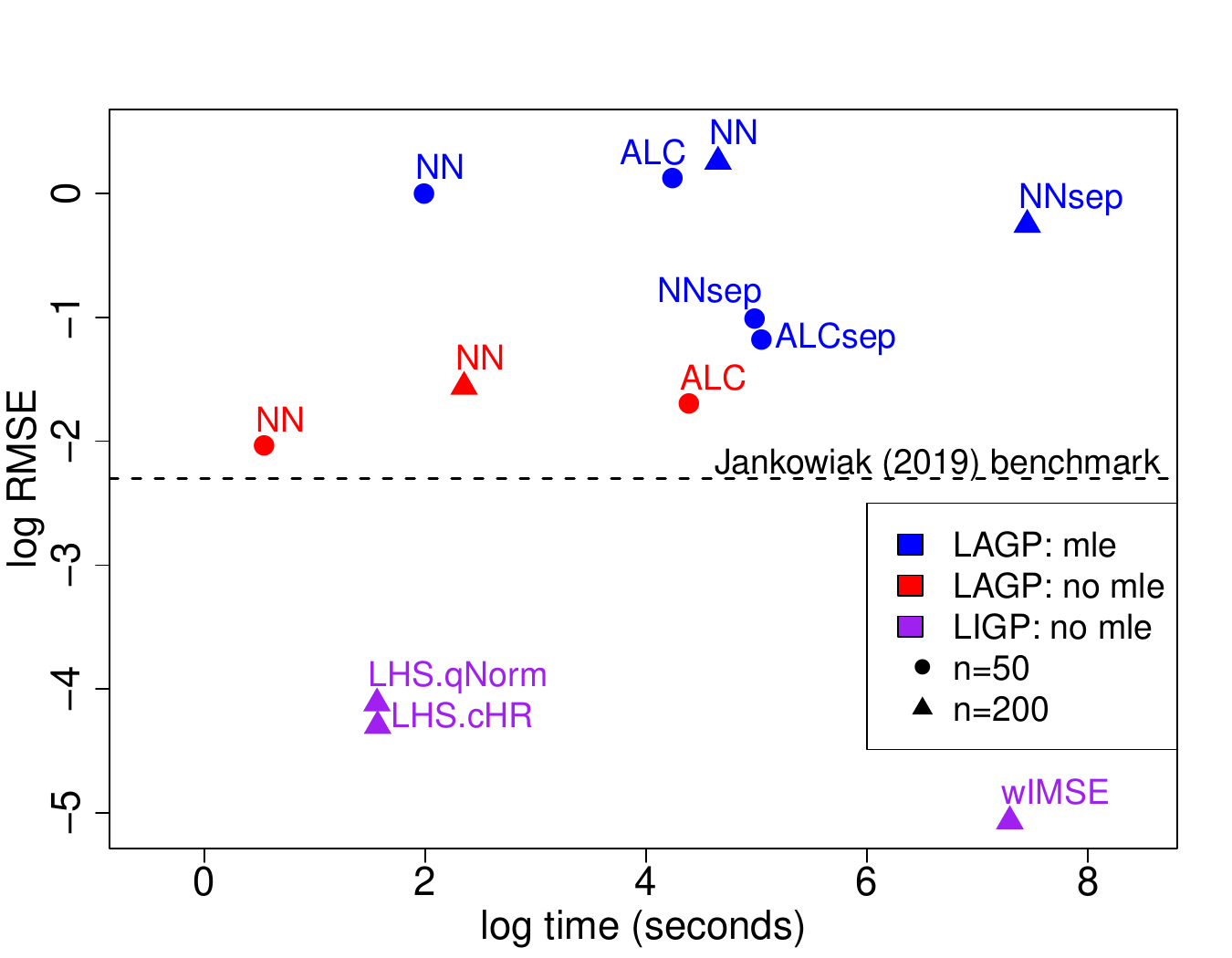}
		\caption{LAGP v.~LIGP models pitting log RMSE ($y$-axis)
			against log time ($x$-axis) on SARCOS data. LAGP fits included both isotropic
			and anisotropic (sep) local lengthscales. Fixing local $\theta_0 = 1$ (no mle) yields computational and predictive advantages.}
		\label{fig:sarcos_map}
	\end{figure}
	
	Figure \ref{fig:sarcos_map} summarizes those results, plotting log RMSE
	against log computation time. Working from the top of the figure
	(lowest predictive accuracy) downwards, observe that default LAGP (blue),
	i.e., with local MLE lenthscales, performs worst.  Larger local neighborhoods
	($n=200$ vs.~$n=50$) do not help accuracy much, and hurt speed.
	Separable lengthscales improve accuracy by an order of magnitude, but you do
	even better by sticking with a fixed $\theta_0=1$ after pre-scaling, which
	brings us to the second (red) group. Foregoing local MLE calculation conveys a
	several orders-of-magnitude speed-up.
	These RMSEs are on par with the best methods in recent studies.  For example,
	\cite{jankowiak2019neural} report on a bakeoff of ten deep and shallow GP and
	neural network comparators, with best RMSE of 0.107, which in log space is
	$-2.3$ (dashed horizontal line).\footnote{No timings provided; the worst
		method had RMSE 0.25.} Keeping it simple in high dimension,
	especially when the training data lie on a lower-dimensional
	manifold, helps control estimation risk and enhances stability.  Larger neighborhoods give a small accuracy fillip,  but
	substantial increase in computation time.
	
	Finally, LIGP methods $(m,n)=(80,200)$ fall into the last/lowest (purple)
	group with the highest accuracy. These are 4-5 orders of magnitude more
	accurate than the default LAGP setup, 2-3 orders better than nomle-LAGP.  Compute times are
	commensurate with the red/middle group, excepting two cases. An wIMSE template
	pays accuracy dividends for increased computational cost.  Simple LAGP (NN) is
	faster but substantially less accurate.  We again remind that these timings
	are unfair to LIGP's {\sf R}-only implementation.

	\subsection{Satellite Drag}
	
	Finally, consider large data sets of simulated drag coefficients for
	satellites in low-Earth orbit. For a description of these data see
	\cite{sun2019emulating}, \cite{mehta2014modeling}, \citet[][Chapter
	2.3.3]{gramacy2020surrogates} and the Git repo
	\url{https://bitbucket.org/gramacylab/tpm/src}. We seek accurate surrogates
	for drag for the Hubble Space Telescope (HST). Simulations, via so-called
	called test particle MC (TPMC), treat atmospheric elements of atomic oxygen
	(O), molecular oxygen ($\text{O}_2$), atomic nitrogen (N), molecular nitrogen
	($\text{N}_2$), helium (He), or hydrogen (H) separately. Following previous
	studies, we consider surrogates for these ``species'' separately.  Data for
	each species is comprised of a two million-sized ($N$) LHS over eight
	configuration inputs.  The goal is to predict drag to a 1\% relative RMSE
	(RMSPE) accuracy. Big training data are essential to meeting that benchmark,
	and needless to say ordinary large-$N$ GP surrogates are not a viable
	alternative.
	
	\begin{figure*}[ht!]
		\centering
		\begin{subfigure}{.96\textwidth}
			\includegraphics[trim= 0 9.8cm 0 8cm, clip, width=\textwidth]{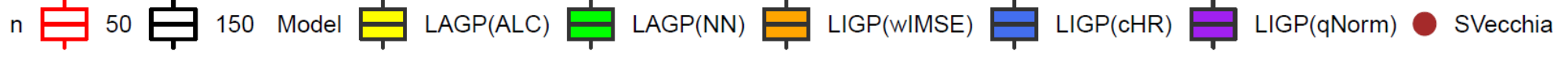}
		\end{subfigure}
		%\begin{subfigure}{.175\textwidth}
		%	\includegraphics[trim=41cm 0 .5cm 15cm, clip, width=\textwidth]{sat_drag_legend_pt1.pdf}
		%\end{subfigure}
		%\begin{subfigure}{.70\textwidth}
		%	\includegraphics[trim=5.5cm 0 9.9cm 15cm, clip, width=\textwidth]{sat_drag_legend_pt1.pdf}
		%\end{subfigure}
	%\begin{subfigure}{.105\textwidth}
	%	\includegraphics[trim=25.5cm 0 20cm 20cm, clip, width=\textwidth]{sat_drag_legend_pt2.pdf}
	%\end{subfigure}
		\begin{subfigure}{0.49\textwidth}
			\includegraphics[trim=0 .5cm 0 0, clip, width=\textwidth]{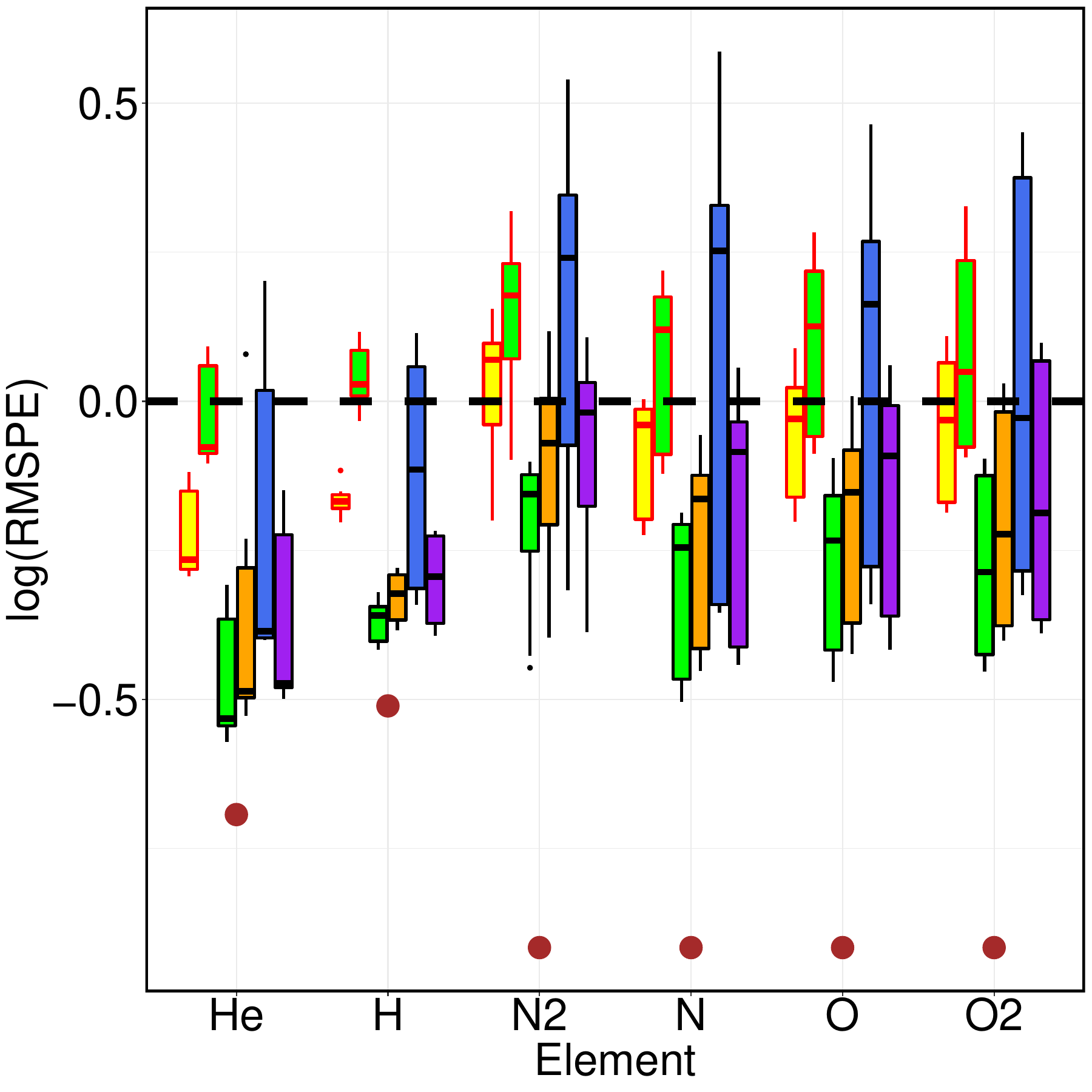}
		\end{subfigure}
		\begin{subfigure}{0.49\textwidth}
			\includegraphics[trim=0 .5cm 0 0, clip, width=\textwidth]{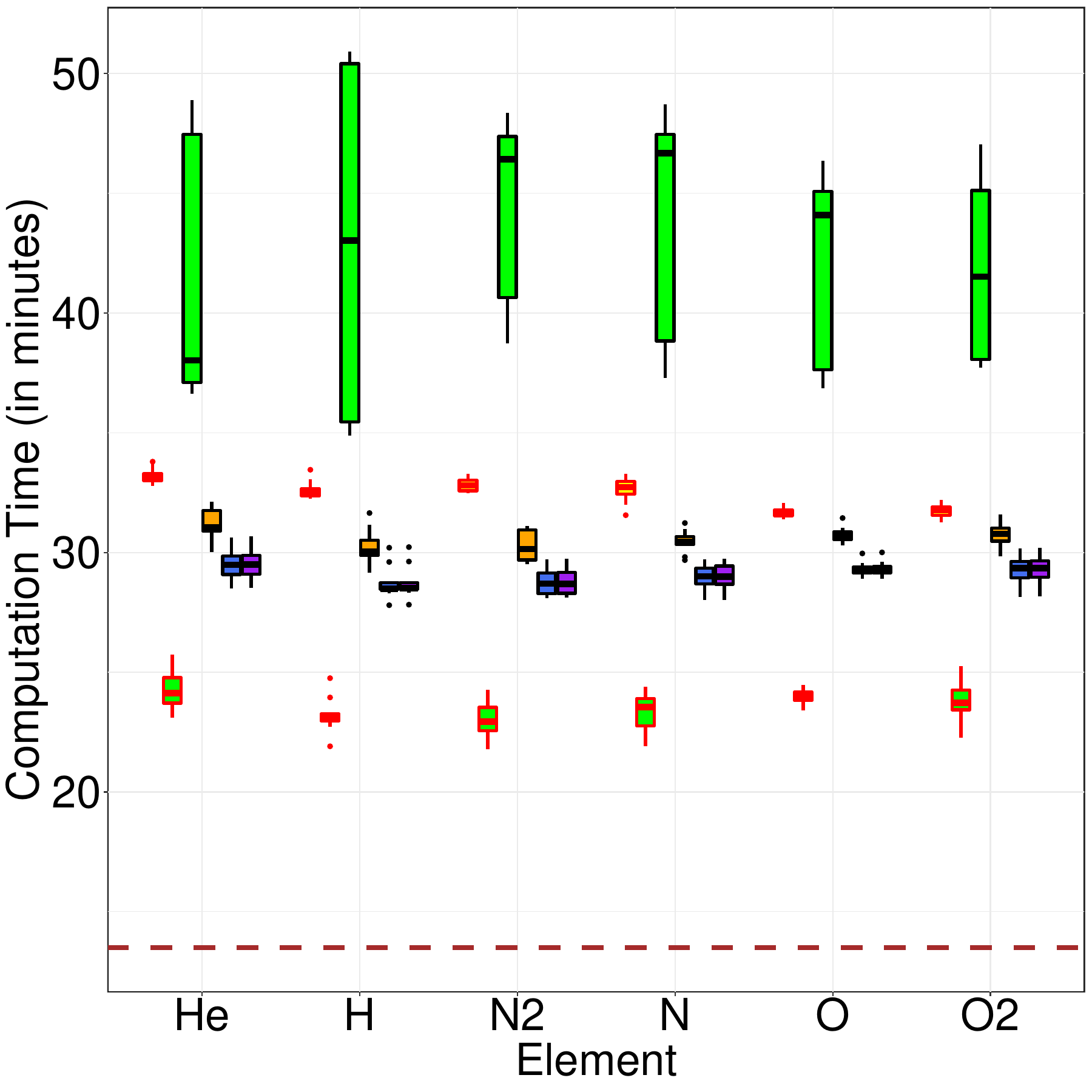} 
		\end{subfigure}
		\caption{{\em Left:} accuracy over 10-fold cross validation for each species
			via log RMSPE.   The horizontal line denotes the 1\% benchmark in log space.
			{\em Right:} Prediction compute time (in minutes) across cross-validation
			folds.}
		\label{fig:satdrag_boxplot}
	\end{figure*}
	
	Figure \ref{fig:satdrag_boxplot} summarizes the results of 10-fold
	cross-validation for each species.  The 1\% benchmark is shown horizontally at
	zero in log space.    Again mimicking previous experiments, we pre-scale
	(Section \ref{sec:implement}) after coding inputs and  before fitting local
	approximations.  Observe in the left panel that LAGP (NN) with $n=150$ is the
	only method able to produce log RMSPEs below the 1\% benchmark for all folds.
	However, LIGP (wIMSE) and LIGP (qNorm) come in at a close second and third
	and have medians (over all folds) below the 1\% benchmark. Factoring in
	computation time (right panel), LIGP methods predict roughly 50\% faster than
	LAGP (NN) with $n=150$. Given the scale of the test and training sets, even
	LIGP (wIMSE) emerges as a viable, cheap alternative.
	
	In contrast to the previous two examples, LIGP (cHR) accuracy suffers
	relative to the other space-filling template scheme LIGP (qNorm). This may
	be due to nonstationarity. Inducing points that lie within the
	neighborhood -- thus motivating LIGP (qNorm) -- transfer more of the
	flexible structure of the GP and provide more accurate predictions.
	Finally, results recently released using SVecchia (brown) offer further
	improvement, although only when substantial training time is amortized
	over a large predictive set. In cases when a single or a relatively small
	number of predictions are needed, LIGP/LAGP can furnish accurate
	predictions in seconds, whereas SVecchia requires (tens of) minutes.
	
	\section{Discussion}
	\label{sec:discuss}
	
	Exponential growth of diversity and size of computer simulation campaigns
	places a heavy burden on GP surrogates.  Remaining fast enough to be useful --
	they cannot be slower than the simulator they are replacing -- but without
	cutting too many corners in approximation, in order to keep fidelity high to
	capture nonstationary relationships, requires a nimble approach.  Many
	interesting new methods have come online of late, including inducing points
	and local approximation.  Inducing points address computation time and space
	head on, but sacrifice on fidelity.  Existing likelihood based tools for choosing their multiplicity and location are difficult to wield due to an abundance of local
	minima. Local approximations (LAGP) perform  better in prediction
	exercises because their criteria more squarely target predictive accuracy. 
	However, they rely on cumbersome discrete search to
	supplant intractably large conditioning sets.  
	
	Here we proposed a hybrid approach: locally induced Gaussian processes
	(LIGPs). Toward that end, we developed a novel weighted integrated
	mean-squared error (wIMSE) criterion for selecting inducing points
	nearby predictive locations of interest.  Closed forms for the criteria and
	derivatives were provided.  The key insight here is one of replacing discrete
	data subset selection (LAGP) with continuous, library-based search via wIMSE
	through inducing points.  Our empirical work revealed that such conditioning
	sets had a highly consistent structure from one predictive location to the
	next, suggesting that one-off calculations could be reused as a template for
	other locations of interest.
	
	The result is a new transductive GP learner that is faster than the
	original, with comparable or improved accuracy in out-of-sample
	exercises.  When LIGP results are less accurate than LAGP, the
	gaps are narrow and LAGP requires substantially more computation. 
	In some cases, LIGP is orders of magnitude more accurate without demanding
	more computation.  Our examples spanned illustrative (2d and 8d with tens
	and hundreds thousands of points) to high-dimensional benchmarks (21d with
	non-space-filling design) and real-world simulation (8d and millions of runs).
	
	We see these promising results as providing a solid foundation from which to
	explore improvements: from accurate and even faster predictions; to broader
	application such as in low-signal and even heteroskedastic \citep{hetGP2}
	stochastic simulation experiments. We have some specific ideas. Rather than NN
	neighborhoods for each predictive location, thrifty ALC alternatives
	\citep[e.g., {\tt alcray} in {\tt laGP},][]{gramacy2016speeding} may enhance
	the hybrid. Kernel support could be expanded to include other families, such
	as Mat\'ern, or to include locally separable lengthscales.  In addition, automating the
	choice of local sizes $(m,n)$ through a Bayesian optimization of out-of-sample RMSE could help make the methodology more
	plug-n-play.

%\begin{acknowledgements}
%	 We would like to thank the journal editor and referees for their thorough review of this paper. They provided valuable insights and suggestions, helping improve the narrative and context of this work. DAC and RBG recognize support from National Science Foundation (NSF) grant DMS-1821258.
%\end{acknowledgements}

	\bibliographystyle{jasa}
	\bibliography{liGP_articles}
	
	\appendix
	\section{IMSE and ALC overview} \label{app:imse}
	As mentioned in Section \ref{sec:opt}, variance-based sequential design criteria are better aligned with the goal of generating accurate GP predictions than using the likelihood. We consider variations on integrated mean-squared
	error (IMSE) over a domain $\mathcal{X}$, with smaller being
	better:
	\[
	\mathrm{(IMSE)} \quad\quad
	I = \int_{\tilde{\x}\in \mathcal{X}}\sigma^2(\tilde{\x}) \; d\tilde{\x}.
	\]
	Choose $\sigma^2(\cdot) \equiv \sigma_N^2(\cdot)/\nu$ from
	Eq.~\eqref{eq:GPgen}, and $I$ may be used to optimize the $N$ coordinates of
	$\X_N$, or to choose the next ($N+1^\mathrm{st}$) one ($\tilde{\x}_{N+1})$ in
	a sequential setting.\footnote{Dividing out $\nu$ removes
		dependence on $\Y$-values through $\hat{\nu}$.  Greedy build-up of $\x_{n+1}$
		over $n=N_0, \dots,N-1$ is near optimal due to a supermartingale property
		\citep{bect2016supermartingale}.}  Closed form expressions are available for
	rectangular $\mathcal{X}$ and common kernels
	\citep[e.g.,][]{ankenman2010,anagnostopoulos2013,burnaev2015,leatherman2018}.
	Analytic derivatives $\frac{\partial I}{\partial \tilde{\x}_{N+1}}$ facilitate
	numerical optimization \citep[][Chapters 4 \&
	10]{hetGP2,gramacy2020surrogates}.  Approximations are common otherwise
	\citep{gramacy2009,gauthier2014,gorodetsky2016,pratola2017}.
	
	An analogue active learning heuristic from
	\cite{Cohn1993}, dubbed ALC, instead targets variance aggregated over a
	discrete reference set $\mathcal{X}$, originally for neural network surrogates:
	\[
	\mathrm{(ALC)} \quad\quad
	\Delta \sigma^2 =\sum_{\tilde{\x}\in \mathcal{X}}\sigma^2(\tilde{\x})-\sigma_{\mathrm{new}}^2(\tilde{\x}),
	\]
	\cite{Seo2000} ported ALC to GPs taking $\sigma^2(\cdot) = \sigma_N^2(\cdot)$
	and $\sigma^2_{\mathrm{new}}(\cdot) \equiv \sigma^2_{N+1}(\cdot)$.  If
	discrete and volume-based $\mathcal{X}$ are similar, then $\Delta \sigma^2
	\approx c
	- I$, where $c$ is constant on $\x_{N+1}$.  Discrete $\Delta \sigma^2$ via ALC is advantageous in transductive learning settings \citep{vapnik2013nature}, where $\mathcal{X}$ can be matched with a testing set.  Otherwise, analytic $I$ via IMSE may be preferred.

	Against that backdrop, we propose employing ALC and IMSE to select inducing
	points $\bar{\X}_M$. To our knowledge, using such variance-based criteria is novel in the literature on the selection of inducing points.  The criteria below are framed sequentially, for an
	$M+1^\mathrm{st}$ point given $M$ collected already.  Although we prefer this
	greedy approach -- optimizing $d$ coordinates one-at-a-time rather than $Md$
	all at once in a surface with many equivalent locally optimal configurations
	due to label-switching -- either criteria is easily re-purposed for an
	all-at-once application.  Under the diagonal-corrected Nystr\"om approximation
	\eqref{eq:SPGP} and assuming coded $\mathcal{X} = [0,1]^d$,
	\begin{align} \label{eq:alcip}
			\text{ALC}_N^{(M+1)} &= \text{ALC}(\bar{\x}_{M+1}; \X_N, \Y_N, \mathcal{X},\bar{\X}_M) =c-\sum_{\tilde{\x}\in \mathcal{X}}\sigma_{M+1,N}^2(\tilde{\x})
			, \quad \mbox{ and } \\
			\text{IMSE}_N^{(M+1)} &= \text{IMSE}(\bar{\x}_{M+1},\X_N, \Y_N, \mathcal{X},\bar{\X}_M)=E-\text{tr}\Big\{\Big(\K_{M+1}^{-1}-\Q_{M+1}^{-1(N)}\Big)\W_{M+1}\Big\}, \nonumber
	\end{align}
	where $E=\int_{\tilde{x}\in \mathcal{X}}k(\tilde{\x},\tilde{\x})d\tilde{\x}$
	and $\W_{M+1}$ is  $(M+1)\times (M+1)$ via $
	w(\bar{\x}_i,\bar{\x}_j)=\int_{\tilde{\x}\in
		\mathcal{X}}k(\bar{\x}_i,\tilde{\x})k(\bar{\x}_j,\tilde{\x})d\tilde{\x}$
		for
	$i,j\in\{1,...,M+1\}$. This derivation is similar to the wIMSE calculations \eqref{eq:wimse} and \eqref{eq:wij} following \cite{hetGP2}.	
	\begin{figure*}[ht!]
		\centering
		\begin{subfigure}{0.47\textwidth}
			\includegraphics[trim= 0 10 38 38,clip, width=\textwidth]{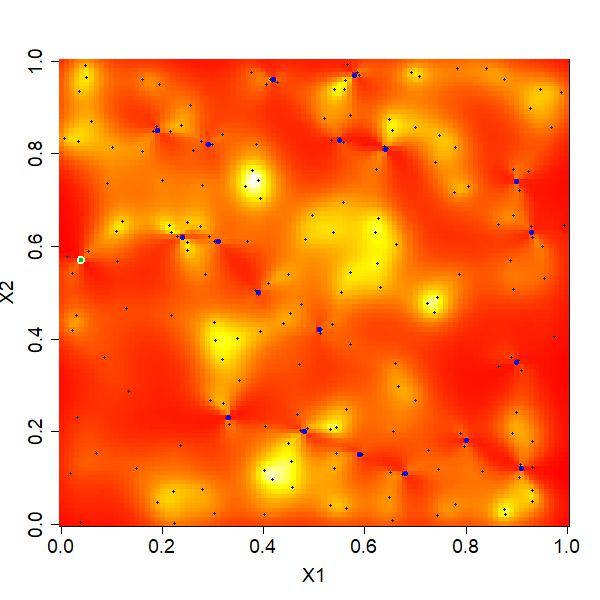}
			\caption{Variational lower bound of log-likelihood surface} \label{fig:optnll}
		\end{subfigure}
		\hspace*{\fill}
		\begin{subfigure}{0.475\textwidth}
			\includegraphics[trim=0 0 38 48, clip, width=\textwidth]{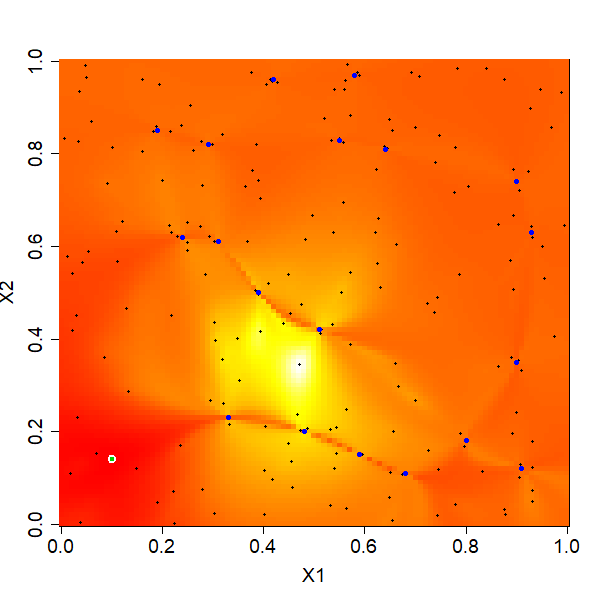}
			\caption{ALC/IMSE surface} \label{fig:optalc}
		\end{subfigure}
		\caption{In both panels: $N=200$ training data points (black dots) and
			$M=19$ inducing points (blue dots), selecting the twentieth one (green)
			by two criteria: (a) variational lower bound of the log-likelihood; (b) ALC/IMSE.  Yellow is higher/red
			lower.}
		\label{fig:optbakeoff}
	\end{figure*}
	
	\sloppy To explore inducing point optimization, consider Herbie's tooth
	\citep{herbtooth} described in Section \ref{sec:opt}. Figure
	\ref{fig:optbakeoff} shows variational lower-bound of the log-likelihood (left) and ALC/IMSE
	surfaces (right) for $\bar{\x}_{20}$ given a modestly sized training
	dataset $(\X_N,\Y_N)$ of size $N=200$. Similarities in the two surfaces
	are apparent.  Many low/red areas coincide, but the optimizing locations
	(green dots), found via multi-start local optimization with identically
	fixed kernel hyperparameters, do not. Even after taking great care to
	humbly restrict searchers, e.g., from crossing $\bar{\X}_M$ locations,
	sometimes upwards of 1000 evaluations were required to achieve
	convergence. Consequently, quadratic ALC/IMSE is faster.
	
	\subsection{Derivations of wIMSE and its gradient}
	\label{app:wimse}

For the predictive location $\x^*$, assign weight
$k_\theta(\tilde{\x},\x^\star)$ and consider squared exponential kernel  $k_\theta(\cdot, \cdot)$ with isotropic lengthscale \eqref{eq:kernel}.  The following is based
on predictive variance \eqref{eq:GPpred} and expectation of
the quadratic form of a random vector \citep[][Section 3.1]{hetGP2}.
		\begin{flalign*}
		& \text{wIMSE}(\bar{\x}_{m+1},\mathcal{X},\bar{\X}_m, \X_n,\theta, \x^\star)  = \int_{\tilde{\x}\in\mathcal{X}}k_\theta(\tilde{\x},\x^\star)\sigma^2_{n,m+1}(\tilde{\x})\hspace{2mm}d\tilde{\x}\\
		& = \int_{\tilde{\x}\in\mathcal{X}}k_\theta(\tilde{\x},\x^\star)\Big(k_\theta(\tilde{\x},\tilde{\x}) + \epsilon_K  - k_\theta(\tilde{\x}, \bar{\X}_{m+1})\Big[\K^{-1}_{m+1}-\Q^{-1(n)}_{m+1}\Big]k_\theta(\tilde{\x}, \bar{\X}_{m+1})^\top\Big) d\tilde{\x}\\
		& = \prod_{k=1}^D \Big((1+\epsilon_K)\int_{a_k}^{b_k}k_\theta(\tilde{\x}_k,\x^\star_k) d\tilde{\x}_k \\
		& \quad - \int_{a_k}^{b_k}k_\theta(\tilde{\x}_k, \x^\star_k)^{1/2}k_\theta(\tilde{\x}_k, \bar{\X}_{m+1,k})\Big[\K^{-1}_{m+1}-\Q^{-1(n)}_{m+1}\Big]\times k_\theta(\tilde{\x}_k, \bar{\X}_{m+1,k})^\top k_\theta(\tilde{\x}_k,\x^\star_k)^{1/2} \hspace{2mm}d\tilde{\x}_k\Big)\\
		& =\frac{\sqrt{\theta\pi}(1+\epsilon_K)^D}{2}\prod_{k=1}^D\left(\text{erf}\left\{\frac{\x^\star-a_k}{\sqrt{\theta}}\right\}-\text{erf}\left\{\frac{\x^\star-b_k}{\sqrt{\theta}}\right\}\right)-\text{tr}\Big\{(\K^{-1}_{m+1}-\Q^{-1}_{m+1})\W^\star_{m+1}\Big\}\\
	\end{flalign*}
	where $\W^\star_{m+1}=\prod_{k=1}^D \W^\star_{m+1,k}$ in $D$ dimensions. The entry in the $i^{\text{th}}$ row and $j^{\text{th}}$ column of $\W^\star_{m+1,k}$ is 
	\begin{flalign*}
		w^{\star(i,j)}_{m+1,k}&\equiv w^\star_{m+1,k}(\bar{\x}_{i,k}, \bar{\x}_{j,k}) \\
		&= \int_{a_k}^{b_k}k_\theta(\tilde{\x}_k,\x^\star_k)k_\theta(\tilde{\x}_k,\bar{\x}_{i,k})k_\theta(\tilde{\x}_k,\bar{\x}_{j,k})d\tilde{\x}_k \\
		&=\int_{a_k}^{b_k}\!\!\!\!\exp\left\{-\frac{(\tilde{\x}_k-\x^\star_k)^2+(\tilde{\x}_k\!-\!\bar{\x}_{i,k})^2+(\tilde{\x}_k\!-\!\bar{\x}_{j,k})^2}{\theta}\right\}d\tilde{\x}_k\\
		&=\sqrt{\frac{\pi\theta}{12}}\exp\Big\{\frac{2}{3\theta}(\bar{\x}_{i,k}\x^\star_k + \bar{\x}_{j,k}\x^\star_k + \bar{\x}_{i,k}\bar{\x}_{j,k} - \x^{*2}_k - \bar{\x}_{i,k}^2 - \bar{\x}_{j,k}^2 )\Big\}\times\\
		&\hspace{20mm}\left(\text{erf}\left\{\frac{\iota^{(i,j)}_{k}-3a_k}{\sqrt{3\theta}}\right\}-\text{erf}\left\{\frac{\iota^{(i,j)}_{k}-3b_k}{\sqrt{3\theta}}\right\}\right)
	\end{flalign*}
	where $\iota^{(i,j)}_{k}=\x^\star_{k}+\bar{\x}_{i,k}+\bar{\x}_{j,k}$. $\bar{\x}_{i,k},\bar{\x}_{j,k}$ are entries from the $i^{\text{th}}$ and $j^{\text{th}}$ rows and $k^{\text{th}}$ column of $\bar{\X}_{m+1}$ ($i, j \in \{1,\dots,m+1\}$) and $\x^\star_k$ is the $k^{\text{th}}$ coordinate of $\x^\star$.
	
	\label{app:wimse_grad}
	The gradient of weighted integrated mean-squared error with respect to the $k^{\text{th}}$ dimension of $\bar{\x}_{m+1}$ is:
		\begin{flalign*}
		&\frac{\partial\text{wIMSE}(\bar{\x}_{m+1},\mathcal{X},\bar{\X}_m, \X_n,\theta, \x^\star)}{\partial\bar{\x}_{m+1,k}} \\
		&\qquad = - \text{tr}\left\{\left(\frac{\partial\K^{-1}_{m+1}}{\partial\bar{\x}_{m+1,k}}-\frac{\partial\Q^{-1(n)}_{m+1}}{\partial\bar{\x}_{m+1,k}}\right)\W^\star_{m+1}\right\} - \text{tr}\left\{\left(\K^{-1}_{m+1}-\Q^{-1(n)}_{m+1}\right)\frac{\partial\W^\star_{m+1}}{\partial\bar{\x}_{m+1,k}}\right\}\\
		&\qquad= \text{tr}\Big\{\Big(\K^{-1}_{m+1}\frac{\partial\K_{m+1}}{\partial\bar{\x}_{m+1,k}}\K^{-1}_{m+1}-\Q^{-1(n)}_{m+1}\frac{\partial\Q^{(n)}_{m+1}}{\partial\bar{\x}_{m+1,k}}\Q^{-1(n)}_{m+1}\Big)\W^\star_{m+1}\Big\}\\
		&\hspace{15mm}- \text{tr}\left\{\left(\K^{-1}_{m+1}-\Q^{-1(n)}_{m+1}\right)\frac{\partial\W^\star_{m+1}}{\partial\bar{\x}_{m+1,k}}\right\}
	\end{flalign*}
	In the matrix $\frac{\partial\W^\star_{m+1}}{d\bar{\x}_{m+1,k}}$, all entries are zero except the row/column that corresponds to the row of $\bar{\X}_{m+1}$ that contains $\bar{\x}_{m+1}$, which we place in the last $m+1^{\text{st}}$ row. For the nonzero entries in $\frac{\partial\W^\star_{m+1}}{\partial\bar{\x}_{m+1,k}}$, we re-express them as
	$$\frac{\partial w^\star_{m+1}(\bar{\x}_i,\bar{\x}_{m+1})}{\partial\bar{\x}_{m+1,k}}=\frac{\partial w^{\star(i,m+1)}_{m+1}}{\partial\bar{\x}_{m+1,k}}\prod_{k'=1,k'\neq k}^D w^{\star(i,m+1)}_{m+1,k'}$$ 
	where
	\begin{flalign*}
		\frac{\partial w^{\star(i,m+1)}_{m+1}}{\partial\bar{\x}_{m+1,k}}&=\sqrt{\frac{\pi\theta}{12}}\text{exp}\Bigg\{\frac{2}{3\theta}\Big(\bar{\x}_{i,k}\x^\star_{k} + \bar{\x}_{m+1,k}\x^\star_{k} + \bar{\x}_{i,k'}\bar{\x}_{m+1,k} - \x^{*2}_{k} - \bar{\x}_{i,k}^2 - \bar{\x}_{m+1,k}^2 \Big)\Bigg\}\\
		&\qquad\times \Bigg[\frac{2}{3\theta}(\x_{k}^\star-2\bar{\x}_{m+1,k}-\bar{\x}_{i,k})\times \left(\text{erf}\left\{\frac{\iota^{(i,m+1)}_{k}-3a_{k}}{\sqrt{3\theta}}\right\}-\text{erf}\left\{\frac{\iota^{(i,m+1)}_{k}-3b_{k}}{\sqrt{3\theta}}\right\}\right)\\
		&\quad\qquad+\frac{2}{\sqrt{3\pi\theta}}\Bigg(\text{exp}\Big\{-\frac{(\iota^{(i,m+1)}_{k}-3a_{k})^2}{3\theta}\Big\}-\text{exp}\Big\{-\frac{(\iota^{(i,m+1)}_{k}-3b_{k})^2}{3\theta}\Big\}\Bigg)\Bigg].
	\end{flalign*}

\label{app:KQpartitions}

Working with
$\K_{m+1}$ and $\Q_{m+1}^{(n)}$ is cubic in $m$, yet even that is
overkill. Thrifty evaluation of
Eqs.~(\ref{eq:wimse}--\ref{eq:dwimse}) lies in construction of
$\Q_{m+1}^{(n)}$ which is equivalent to $\K_{m+1} +
\kk^\top_{n,m+1}\mathbf{\Gamma}_{n,m+1}$. Evaluating
$\kk^\top_{n,m+1}\mathbf{\Gamma}_{n,m+1}$ requires $2n-1$ products for
each of $(m+1)^2$ entries, incurring costs in $\mathcal{O}(m^2n)$ flops.
Assuming $n\gg m$, this dominates the $\OO(m^3)$ cost of decomposition.

More time can be saved through partitioned inverse \citep{barnett:1979}
sequential updates to $\K^{-1}_{m+1}$ after the new $\bar{\x}_{m+1}$ is
chosen, porting LAGPs frugal updates to the LIGP context. Writing $\K_{m+1}$ as an
$m$-submatrix with new $m+1^\mathrm{st}$ column gives
\begin{align}  \label{eq:ki_seq}
	%\begin{split}
		\K_{m+1}=\begin{bmatrix} \K_m & \kk_m(\bar{\x}_{m+1}) \\ 
			\kk_m(\bar{\x}_{m+1})^\top & k_\theta(\bar{\x}_{m+1},\bar{\x}_{m+1}) \end{bmatrix}
		 \mbox{ so that }
		\K_{m+1}^{-1}= \begin{bmatrix}
			\K_m^{-1}+\rho\boldsymbol{\eta} 
			\boldsymbol{\eta}^\top 
			& \boldsymbol{\eta} \\ 
			\boldsymbol{\eta}^\top 
			& \rho^{-1}&
		\end{bmatrix}
	%\end{split}
\end{align}
using $\rho = k_\theta(\bar{\x}_{m+1},\bar{\x}_{m+1})
-\kk_m^\top(\bar{\x}_{m+1})\K_m^{-1}\kk_m(\bar{\x}_{m+1})$ and $m$-length
column vector $\boldsymbol{\eta}=-\rho^{-1}\K_m^{-1}\kk_m(\bar{\x}_{m+1})$.
Updating $\K_{m+1}^{-1}$ requires calculation of $\rho$,
$\boldsymbol{\eta}$, and $\boldsymbol{\eta}
\boldsymbol{\eta}^\top$, each of which is in
$\mathcal{O}(m^2)$. Thus we reduce the computational complexity of
$\K_{m+1}^{-1}$ from $\OO\left(m^3\right)$ to $\OO\left(m^2\right)$.  Similar
partitioning provides sequential updates to $\Omega_n^{(m+1)}$, a diagonal
matrix:
\begin{align} 
	\Omega_n^{(m+1)}&=\text{Diag}\!\left(\K_{n}+\epsilon_K\I_n-\kk_{n,m+1}\K_{m+1}^{-1} \kk_{n,m+1}^\top\!\right) \nonumber\\
	&=\Omega_n^{(m)}-\rho^{-1}\text{Diag}\left\{\zeta\zeta^\top\right\} \label{eq:Om_seq}
\end{align}
where $\zeta=\kk_{nm}\K_m^{-1}\kk_m(\bar{\x}_{m+1})-\kk_n(\bar{\x}_{m+1})$.
Updates of $\Omega_n^{(m+1)}$ without partitioning, driven by matrix--vector
product(s) $\kk_{n,m+1}\K_{m+1}^{-1} \kk_{n,m+1}^\top$ involve $m^2n$
flops. Using
\eqref{eq:Om_seq} reduces that to $\mathcal{O}(mn)$.

Unlike in Eq.~\eqref{eq:ki_seq}, $\Q_{m}^{(n)}$ cannot be trivially augmented
to construct $\Q_{m+1}^{(n)}$ due to the presence of $\Omega_n^{(m)}$ which
is also embedded in $\Q_{m}^{(n)}$. Yet there are some time savings to be found
in the partitioned inverse
\begin{align} \label{eq:Q_seq}
	%\begin{split}
		\Q_{m+1}^{(n)}=\begin{bmatrix}  \Q_{m*}^{(n)} & \boldsymbol{\gamma}(\bar{\x}_{m+1}) \\ \boldsymbol{\gamma}(\bar{\x}_{m+1})^\top & \psi(\bar{\x}_{m+1}) \end{bmatrix}\quad
		\Q_{m+1}^{-1(n)}
		=\begin{bmatrix} \Q_{m*}^{-1(n)}+\upsilon\boldsymbol{\xi} \boldsymbol{\xi}^\top & 
			\boldsymbol{\xi}\\ 
			\boldsymbol{\xi}^\top & \upsilon^{-1} \end{bmatrix}
	%\end{split}
\end{align}
with $\Q_{m*}^{(n)}= \K_m + \kk_{nm}^\top\Omega_n^{(m+1)-1} \kk_{nm}$ built via updated values of $\Omega_n^{(m+1)}$,
$\boldsymbol{\gamma}(\bar{\x}_{m+1}) =
\kk_m(\bar{\x}_{m+1})+\kk^\top_{nm}\Omega_n^{-1(m+1)} \kk_n(\bar{\x}_{m+1})$,
$\psi(\bar{\x}_{m+1}) =
k_\theta(\bar{\x}_{m+1},\bar{\x}_{m+1})+k_n(\bar{\x}_{m+1})^\top\Omega_n^{-1(m+1)}k_n(\bar{\x}_{m+1})$,
$\upsilon =
\psi(\bar{\x}_{m+1})-\boldsymbol{\gamma}(\bar{\x}_{m+1})^\top
\Q_{m*}^{-1(n)}\boldsymbol{\gamma}(\bar{\x}_{m+1})$ and $\xi=
-\upsilon^{-1}\Q_{m*}^{-1(n)}\boldsymbol{\gamma}(\bar{\x}_{m+1})$.
Similar to $\Q_{m}^{(n)}$, calculating $\Q_{m*}^{(n)}$ requires in flops in 
$\mathcal{O}(m^2n)$.  Consequently the entire scheme can be managed in $\mathcal{O}(m^2n)$.

\section{Determining neighborhood size}
\label{app:nsize}

Little attention is paid in the literature to the choosing the number of
(global) inducing points \citep{Seeger2003,Titsias2009,Azzimonti2016} relative
to problem size $(N,d)$, except on computational grounds -- smaller $M$ is
better. The same is true for local neighborhood size $n$ in LAGP.  Although
there is evidence that the {\tt laGP} default of $n=50$ is too small
\citep{laGP}, especially with larger input dimension $d$, cubically growing
expense in $n$ limits the efficacy of larger $n$ in practice. With local
inducing points this is mitigated through cubic-in-$m$ proxies, allowing
larger local neighborhoods, thus implying more latitude to explore/choose good
$(m,n)$ combinations.  

\begin{figure}[ht!]
	\centering
	\includegraphics[trim=0 18 20 45, clip, scale=0.45]{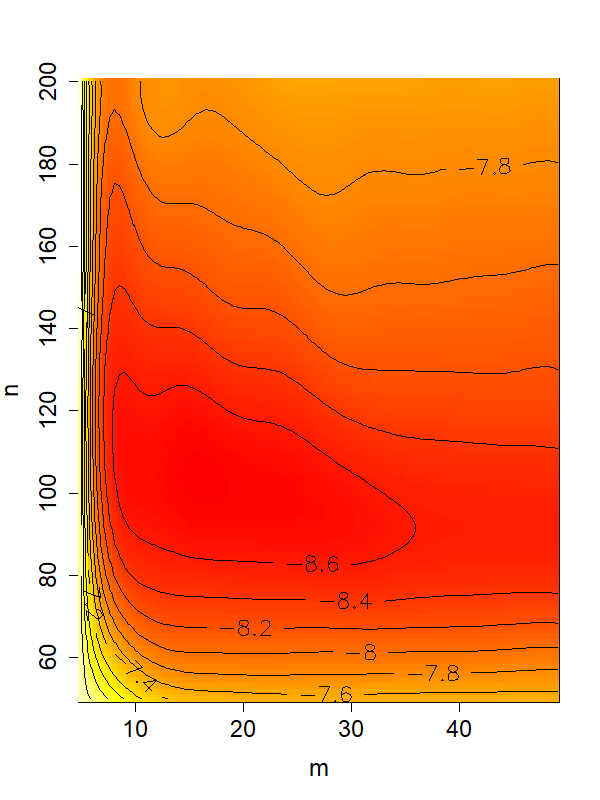}
	\includegraphics[trim=43 18 20 45, clip, scale=0.45]{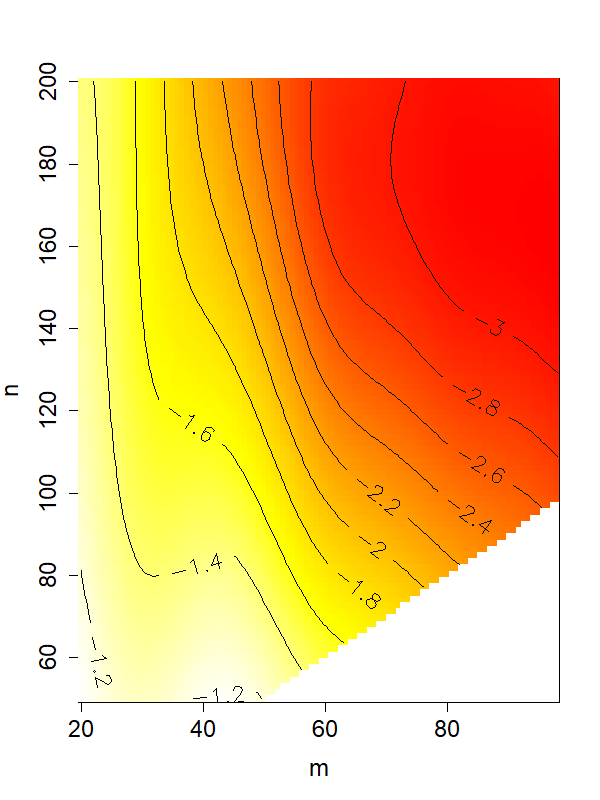}
	\caption{$\log(\mathrm{RMSE})$ over inducing points $m$ and neighborhood $n$: Herbie's tooth (left) and borehole (right).}
	\label{fig:HT_Simulation}
\end{figure}

Toward that end, we considered a coarse grid of $(m,n)$ and predictive RMSEs
on Herbie's tooth ($d=2$) and borehole $(d=8)$ toy problems.  Setup details
are identical to descriptions in Sections \ref{sec:illust} and
\ref{sec:borehole}, respectively, and we used the qNorm ($\Phi^{-1}$) template
throughout. An LHS testing set of size $N'=1000$ was used to generate the
response surfaces of RMSEs reported in Figure \ref{fig:HT_Simulation}. These
are shown in log space for a more visually appealing color scheme, and were
obtained after GP smoothing to remove any artifacts from random testing. Grid
elements where $m>n$ were omitted from the simulation on the grounds that
there are no run-time benefits to those choices.

Observe that both surfaces are fairly flat across a wide swath of $m$,
excepting quick ascent (decrease in accuracy) for smaller numbers of inducing
points in the left panel.  The situation is similar for $n$.  Best settings
are apparently input-dimension dependent.  Numbers of inducing points as low
as $m=10$ seems sufficient in 2d (top panel), whereas $m=80$ is needed in 8d.
For borehole, it appears that larger neighborhoods $n$ are better, perhaps
because the response surface is very smooth and the likelihood prefers long
lengthscales \citep{laGP}.  A setting like $n=150$ seems to offer good results
without being too large.  The situation is different for Herbie's tooth.  Here larger $n$ has deleterious effects.  Its non-stationary nature
demands reactivity which is proffered by smaller local neighborhood.  A
setting of $n=100$ looks good.

These are just two problems, and it is clearly not reasonable to grid-out
$(m,n)$ space for all future applications.  But nevertheless we have found
that these rules of thumb port well to our empirical work in Section
\ref{sec:examples}.  Our satdrag example ($d=8$) and classic
$d=21$ benchmark work well with the settings found for borehole, for example.
Some ideas for automating the choice of $(m,n)$ are discussed in Section
\ref{sec:discuss}.

\end{document}